\documentclass[reprint,aps,prc,amsmath,amssymb,
nofootinbib,
superscriptaddress,
showpacs]{revtex4-1}
\usepackage{graphicx}
\usepackage{dcolumn}
\usepackage{bm}
\usepackage[utf8]{inputenc}

\allowdisplaybreaks[3]
\begin{document}

\title {Methods for analysis of two-particle rapidity correlation function in high-energy heavy-ion collisions}

\author{Ronghua He}
\email{ronghuahe2007@163.com}
\affiliation{Department of Physics, Harbin Institute of Technology, Harbin 150001, People's Republic of China}
\author{Jing Qian}
\affiliation{Department of Physics, Harbin Institute of Technology, Harbin 150001, People's Republic of China}
\author{Lei Huo}
\affiliation{Department of Physics, Harbin Institute of Technology, Harbin 150001, People's Republic of China}

\date{\today}

\begin{abstract}
Two-particle rapidity (or pseudorapidity) correlation function $C(y_1, y_2)$ was used in analysing fluctuation of particle density distribution in rapidity in high-energy heavy-ion collisions. In our research, we argue that for a centrality window, some additional correlation may be caused by a centrality span, when the mean two- and single-particle densities over a centrality window are used directly in the calculation , just like $\left<N(y_1, y_2) \right> / \left[\left<N(y_1)\right>\left<N(y_2)\right>\right]$. We concentrate on removing the influence of collision-centrality span on correlation function, and two calculation methods are raised. In one method, correlation coefficients are considered to be the ratios of probabilities (not the particle density). In the other method, a relative multiplicity is introduced to unity the events of different centralities. For testing the methods, {\sc ampt} model is used and a toy granular model is built to simulate the fluctuation of particle density in rapidity.
\end{abstract}

\maketitle

\section{Introduction\label{sec:introduction}}
\begin{figure}[!t]
\includegraphics[width=8.6cm]{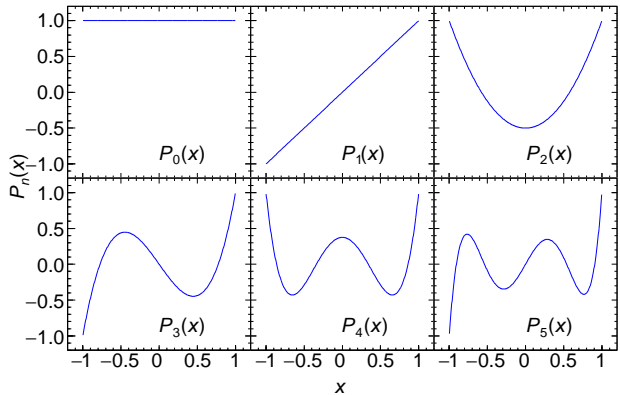}
\caption{\label{fig:illLeg}
Illustration of Legendre polynomials.}
\end{figure}

Hot and dense matter is created in high-energy heavy-ion collisions. The fluctuation of stopping at early stage may lead an event-by-event fluctuation of particle density in rapidity at final stage \cite{2013prcc2, 2016prcc2, 2016prcc2short, 2016cite12, 2016cite14, 2016cite15, 2016cite18, 2016cite19}. The particle density fluctuation in rapidity can be fully parameterized with a group of polynomials \cite{2013prcc2}, such as Chebyshev polynomials \cite{2013prcc2}, Legendre polynomials \cite{2016prcc2} or some other orthogonal polynomials, just like
\begin{equation}
\rho\left(y; a_1, a_2, \cdots\right) = \rho_0\left(y\right) \Big[1 + \sum_{n = 0}^\infty a_n T_n\left(y\right)\Big],
\end{equation}
where $\rho(y; a_1, a_2, \cdots)$ is the single particle density of an event, $\rho_0(y)$ is the mean particle density over the events. In Ref.~\cite{2016prcc2}, $T_n(y) \equiv \sqrt{n+\frac{1}{2}}P_n(y/Y)$, where the $P_0(x) = 1$, $P_1(x) = x$, $P_2(x) = (3x^2-1)/2$, $\cdots$ are Legendre polynomials and illustrated in Fig.~\ref{fig:illLeg}, and $Y$ characterizes the range of rapidity fluctuation, $\left| y \right| < Y$ (Here, $Y = 6$). The event-by-event fluctuation of the rapidity distribution is described by the parameters $a_n$. For instance, $a_1$ reflects the asymmetry \cite{asymmetry1} of particle density distribution in rapidity of an event. $\left<a_1\right>$ is null in the symmetrical collision system, and $\left<a_1^2\right>$ does not have to be null. The variance reflects how the longitudinal shape is changing event by event. On the other hand, purely statistical particle multiplicity fluctuation can also cause event-by-event fluctuation in rapidity. Therefore, $\left<a^2_n\right> = \left<(a_n^\textmd{obs})^2\right> - \left<(a_n^\textmd{ran})^2\right>$ was used for describing pure event-by-event shape fluctuation (nonstatistical component of these longitudinal harmonics) \cite{2016prcc2}, where $a_n^\textmd{obs}$ characterizes the fluctuations due to finite number of particles in the events, and $a_n^\textmd{ran}$ corresponds purely statistical fluctuation, see Ref.~\cite{2016prcc2} or Appendix~\ref{app:aman}.

Two-particle rapidity correlation function $C(y_1, y_2)$ was used for calculating $\left<a_n^2\right>$ \cite{2013prcc2, thesis-c-expPbPb}. The definition of two-particle rapidity correlation function (2PC) can be obtained by comparing the double charged hadrons inclusive cross section $d^2 \sigma_{2\textmd{ch}} / dy_1 dy_2$ to the product of single charged particle inclusive cross sections $d\sigma_{1\textmd{ch}} / dy$. The definition of correlation function $C(y_1, y_2)$ can be expressed as \footnote{The definition of 2PC is borrowed from two-identical-pion correlation function about pion interferometry in Ref.~\cite{hbtprc1979}}
\begin{equation}
C(y_1, y_2) =
\frac
{\left< n_\textmd{ch} \right>^2 }
{\left< n_\textmd{ch} (n_\textmd{ch} - 1)\right>}
\cdot
\frac
{\sigma_\textmd{ch}  d^2\sigma_\textmd{2ch} / dy_1 dy_2}
{ \left( d \sigma_\textmd{1ch} / dy_1 \right) \left( d\sigma_\textmd{1ch} / dy_2 \right) }
\label{equ:definitionOfC}
\end{equation}
where $\sigma_\textmd{ch}$ is the total charged hadron production cross section, and $n_\textmd{ch}$ is the charged hadron multiplicity. The ratio of multiplicity moments is caused by the normalization of single- and double-charged hadron inclusive cross sections as
\begin{eqnarray}
\int dy_1 dy_2 \left(d^2\sigma_\textmd{2ch} / dy_1 dy_2 \right)
&=& \left<n_\textmd{ch} \left(n_\textmd{ch} - 1\right)\right> \sigma_\textmd{ch},\\
\int dy \left( d\sigma_\textmd{1ch} / dy\right) &=& \left<n_\textmd{ch}\right> \sigma_\textmd{ch}.
\end{eqnarray}
This definition Eq.~(\ref{equ:definitionOfC}) insures that $C\left(y_1, y_2\right) \equiv 1$ when the charged hadrons are uncorrelated \cite{hbtprc1979}.

In recent years, two-particle correlation function was usually written as
\begin{equation}
C\left(y_1, y_2\right)
= \frac{\rho_2\left(y_1, y_2\right)}{\rho_1\left(y_1\right)\rho_1\left(y_2\right)},
\label{equ:CRho}
\end{equation}
where $\rho_2(y_1, y_2)$ and $\rho_1(y)$ are two- and single-particle density distribution, respectively \cite{added4, CorrFuncVV2015, CorrFuncPLB2016, CorrFuncPRC2002, CorrFuncJHEP2010, vv2014}. (The differences between Eqs.~(\ref{equ:definitionOfC}) and (\ref{equ:CRho}) are discussed in Sec.~\ref{sec:additionalC} and Appendix~\ref{app:additionalC}.) At a given $a_0$, $a_1$, $\cdots$
\begin{equation}
\rho_2(y_1, y_2; a_0, a_1, \cdots)\! = \!\rho_1(y_1; a_0, a_1, \cdots)\rho_1(y_2; a_0, a_1, \cdots).
\end{equation}
Because $\rho_2(y_1, y_2)$ and $\rho_1(y)$ are averages over $a_i$, 2PC could be expressed as  \cite{2013prcc2, 2016prcc2}
\begin{equation}
C(y_1, y_2)
\!=\! 1\! +\!\!\! \sum_{m, n = 0}^{\infty}\!\!\!\left<a_m a_n\right>
\frac{T_m(y_1)T_n(y_2)+T_m(y_2)T_n(y_1)}{2},
\end{equation}
and $\left<a_m a_n\right>$ was expressed as \cite{2013prcc2, 2016prcc2}
\begin{equation}
\begin{aligned}
\left<a_m a_n\right> =& \frac{1}{Y^2}\int\left[C\left(y_1, y_2\right) - 1\right]\\
&\times \frac{T_m(y_1)T_n(y_2)+T_m(y_2)T_n(y_1)}{2}~dy_1 dy_2,
\end{aligned}
\label{equ:aman}
\end{equation}
see Ref.~\cite{2013prcc2, 2016prcc2} or Appendix~\ref{app:amanFromC}.

But for a centrality window (which is not narrow enough), if the averages $\rho_2(y_1, y_2)$, $\rho_1(y_1)$, and $\rho_1(y_2)$ are calculated over the events of different centralities, we found that some positive additional correlations may be caused (discussed in detail in Sec.~\ref{sec:additionalC} and Appendix~\ref{app:additionalC}). For solving such a problem in the calculation of $C(y_1, y_2)$ for a centrality window, in Ref.~\cite{2016prcc2}, events were divided into narrow centrality intervals according to their total multiplicity. And then 2PCs were calculated in each event class. At last, an average of the 2PCs of the centrality intervals was made as the final result of the whole window. By the way, this method is denoted by C$_\textmd{average}$ for being distinguished with other methods in this paper.

In this paper, we try to calculate $C(y_1, y_2)$ of a centrality window from other ways to remove or reduce the influence of the centrality window span (or the influence of the mixing of events of different centralities), and a multi-phase transport ({\sc ampt}) model \cite{ampt} at $\sqrt{s_{NN}} = 200~\mathrm{GeV}$ of Au+Au collisions is utilized. For testing the methods, we made a toy granular model to simulate the event-by-event fluctuation of particle density in rapidity.

This paper is organized as follows. We discuss the additional positive correlations calculated with Eq.~(\ref{equ:CRho}) in Sec.~\ref{sec:additionalC}. Sec.~\ref{sec:methods} shows the methods used for removing or reducing the influence of centrality span. In Sec.~\ref{sec:granularModel}, the calculation methods are tested with {\sc ampt} model and a toy granular model. A summary is given in Sec.~\ref{sec:summary}.

\section{additional positive correlation\label{sec:additionalC}}

\begin{figure*}[!hbt]
\includegraphics[width=17.8cm]{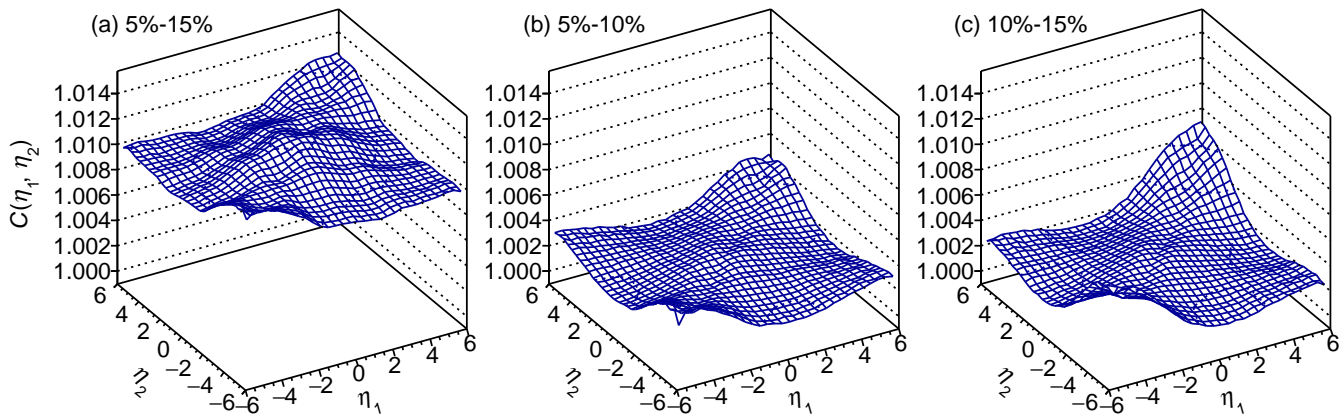}
\caption{\label{fig:ampt}Two-particle pseudorapidity correlation function $C(\eta_1, \eta_2)$ of the string melting version {\sc ampt} model of Au+Au collisions at $\sqrt{s_{NN}} = 200~\mathrm{GeV}$ (partonic scattering section is $3~\mathrm{mb}$), and the results of centrality windows 5\%-15\%, 5\%-10\%, and 10\%-15\% are shown in subgraphs (a), (b), and (c), respectively. }
\end{figure*}

We argue that the mixing of events of different centralities may cause some additional positive correlations, and the reasons are shown as follows.

It is said in section~\ref{sec:introduction} that $\rho_1(y)$ and $\rho_2(y_1, y_2)$ are mean single particle density in rapidity and mean double particle density over events, respectively. If the events are divided into event classes according to the centrality \footnote{Centrality is denoted by $\varepsilon$, and it is worth to note that no matter whether the centrality $\varepsilon$ is decided with multiplicity or some other variables, the discussion in this section will not be influenced.}, single- and double-particle densities in rapidity can be expressed as the averages over the event classes, just like
\begin{equation}
\begin{aligned}
\rho_1(y) &= \left< \rho_1(y; \varepsilon)\right>,\\
\rho_2(y_1, y_2) &= \left< \rho_2(y_1, y_2; \varepsilon)\right>,
\end{aligned}
\end{equation}
where $\rho_1(y;\varepsilon)$ and $\rho_2(y_1, y_2;\varepsilon)$ stand for the mean single- and double-particle densities of an event class around centrality $\varepsilon$. For explaining where the possible additional positive correlation come from, we assume that $C(y_1, y_2; \varepsilon)$ of the event classes are equal to each other (Here, a centrality window divided into $n$ event classes), and $\rho_1(y; \varepsilon)$ decreases with increasing centrality $\varepsilon$. Under these assumptions, 2PC of the whole centrality window is bigger than 2PCs of event classes, just like
\begin{equation}
C\left(y_1, y_2\right) >  \left<C\left(y_1,y_2;\varepsilon\right) \right>= C\left(y_1,y_2;\varepsilon\right),
\end{equation}
because of
\begin{equation}
\frac{\left<\rho_2(y_1,y_2;\varepsilon)\right>}
{\left<\rho_1(y_1;\varepsilon)\right>\left<\rho_1(y_2;\varepsilon)\right>}
>
\frac{\left<\rho_2(y_1,y_2;\varepsilon)\right>}
{\left<\rho_1(y_1;\varepsilon)\rho_1(y_2;\varepsilon)\right>}.
\label{equ:bigger}
\end{equation}
Inequality~(\ref{equ:bigger}) is discussed in detail in Appendix~\ref{app:additionalC}.
We argue that the "additional correlation" is caused by averaging single- and double-particle densities over a centrality window which is not narrow enough.

For describing such a phenomenon, two-particle pseudorapidity correlation function of the string melting {\sc ampt} model (partonic scattering section is $3~\mathrm{mb}$) of Au+Au collisions at $\sqrt{s_{NN}} = 200~\mathrm{GeV}$ is utilized. As shown in Fig.~\ref{fig:ampt}, the centrality window 5\%-15\% is divided into two sub-windows 5\%-10\% and 10\%-15\%, and 2PC of 5\%-15\% (Fig.~\ref{fig:ampt}a) is higher obviously than 2PCs of 5\%-10\% (Fig.~\ref{fig:ampt}b) and 10\%-15\% (Fig.~\ref{fig:ampt}c). We guess that if the influence of centrality window span on $C(\eta_1, \eta_2)$ is removed, 2PC of 5\%-15\% should be in the middle of 2PCs of 5\%-10\% and 10\%-15\% (discussed in Sec.~\ref{sec:granularModel}).

It is worth to note that in the calculation of 2PC in Fig.~\ref{fig:ampt}, $\rho_1(\eta)$ is calculated as the ratio of charged particles number $N(\eta)$ in a pseudorapidity interval around $\eta$ to the interval width $\delta\eta$ (in this paper, $\delta\eta = 0.1$), just like $\left<N(\eta)\right>/\delta\eta$. And similarly,  $\rho_2(\eta_1, \eta_2)$ is calculated with $\left<N(\eta_1) N(\eta_2)\right>/{\delta\eta}^2$. Therefore, $C(\eta_1, \eta_2)$ can be calculated with the equation $\frac{\left<N(\eta_1) N(\eta_2)\right>}{\left<N(\eta_1)\right>\left<N(\eta_2)\right>}$. Considering the situation $\eta_1 = \eta_2$, the expression was written as $\frac{\left<N(\eta_1) N(\eta_2)\right>}{\left<N(\eta_1)\right>\left<N(\eta_2)\right>} - \frac{\delta(\eta_1 - \eta_2)}{\left<N(\eta_1)\right>}$ in Ref.~\cite{2016prcc2}.
On the other hand, for explaining the additional positive correlations, we make an extreme example as follows. We assume that there is no correlation between particles. In other words, particles emit independently and randomly. For a certain centrality $\varepsilon$ (not a centrality window), $C(y_1, y_2; \varepsilon) = \frac{\rho_2(y_1, y_2; \varepsilon)}{\rho_1(y_1; \varepsilon) \rho_1(y_2; \varepsilon)} = 1$, which is equivalent to $\rho_2(y_1, y_2; \varepsilon) = \rho_1(y_1; \varepsilon) \rho_1(y_2; \varepsilon)$. For a centrality window, we assume that $\rho_1(y_1; \varepsilon)$ and $\rho_1(y_2; \varepsilon)$ both decrease with increasing $\varepsilon$, so for the centrality window,
\begin{equation}
\begin{aligned}
\rho_2(y_1, y_2)
&= \left<\rho_2(y_1, y_2; \varepsilon)\right>\\
&=\left<\rho_1(y_1; \varepsilon) \rho_1(y_2; \varepsilon)\right> \\
&> \left<\rho_1(y_1; \varepsilon) \right>\left<\rho_1(y_2; \varepsilon)\right>\\
&= \rho_1(y_1) \rho_1(y_2),
\end{aligned}
\end{equation}
which is equivalent to $C(y_1, y_2)$ bigger than 1. The detailed discussion is shown in Appendix~\ref{app:additionalC}.

Where is the additional positive correlation come from? In the normalized definition of two-particle rapidity correlation function Eq.~(\ref{equ:definitionOfC}), it is obvious when the charged hadrons are uncorrelated, $C(y_1, y_2) \equiv 1$. We guess that the additional positive correlation is caused by the 2PC expression Eq.~(\ref{equ:CRho}) where the particle density distributions are utilized. Single- and double-particle density distributions in rapidity can be expressed as
\begin{eqnarray}
&\rho_1(y_1) =  d\sigma_\textmd{1ch}/ \left(\sigma_\textmd{ch}dy_1\right),~~
\rho_1(y_2) =  d\sigma_\textmd{1ch}/ \left(\sigma_\textmd{ch}dy_2\right),\nonumber\\
&\rho_2(y_1, y_2) = d^2\sigma_\textmd{2ch}/\left(\sigma_\textmd{ch}dy_1 dy_2\right),
\end{eqnarray}
Therefore, the difference between Eqs.~(\ref{equ:definitionOfC}) and (\ref{equ:CRho}) is a ratio $\left<n_\textmd{ch}\right>^2 / \left<n_\textmd{ch}\left(n_\textmd{ch} - 1\right)\right>$. It's notable that for a specific centrality (or a centrality window narrow enough), if we assume that $n_\textmd{ch}$ obeys a Poisson distribution (the variance of $n_\textmd{ch}$ is equal to its expectation), $\left<n_\textmd{ch}\right>^2 / \left<n_\textmd{ch}\left(n_\textmd{ch} - 1\right)\right> = 1$. But when the centrality window is not narrow enough, the assumption of the Poisson distribution may be not suitable \footnote{Some discussions about the influence of centrality window on forward-backward multiplicity correlation strength \cite{rhhe2016, 2009bOfCentralityAndAAModel, FB2017} with a negative binomial distribution are in Ref.~\cite{2008bOfEmitRandomly}.}, and some positive correlation may be added.

\section{methods\label{sec:methods}}

For removing or reducing the influence of the mixing of events of different centralities, in Ref.~\cite{2016prcc2}, events in a centrality window were divided into some narrow centrality intervals. In this section, we try to solve this problem from other ways.

\subsection{C$_\textmd{defintion}$ method}

In this method, we deduce the expression of $C(y_1, y_2)$ (which can be used directly in calculation) from its definition. Eq.~(\ref{equ:definitionOfC}) can be written in another way as
\footnote{The core idea of dealing with two-particle correlation function as the ratio of $P_\textmd{same}(y_1, y_2)$ to $P_\textmd{diff}(y_1, y_2)$ is borrowed from the research about the two-identical-pion interferometry \cite{hbt1, hbt2, hbt3}.}
\begin{eqnarray}
C(y_1, y_2) \!&=&\!
\left.
\frac{d^2\sigma_\textmd{2ch} / dy_1 dy_2}
{\sigma_\textmd{ch}\left< n_\textmd{ch} (n_\textmd{ch} \!-\! 1)\right>}
\middle/\left(\!\frac{d \sigma_\textmd{1ch} / dy_1 }
{\sigma_\textmd{ch}\left< n_\textmd{ch} \right>}
\!\cdot\!\frac{d \sigma_\textmd{1ch} / dy_2 }
{\sigma_\textmd{ch}\left< n_\textmd{ch} \right>}\!\right)
\right.\nonumber\\
&=& \left.P_\textmd{same}\left(y_1, y_2\right)~\middle/~
\left[P_1\left(y_1\right) P_1\left(y_2\right)\right]\right.\nonumber\\
&=& \left.P_\textmd{same}\left(y_1, y_2\right)~\middle/~ P_\textmd{diff}\left(y_1, y_2\right) \right.,
\label{equ:definitionOfC2}
\end{eqnarray}
where $P_\textmd{same}(y_1, y_2)$ is the probability density of a pair of particles with rapidities $y_1$ and $y_2$ chosen from the same event, $P_\textmd{diff}(y_1, y_2)$ is the probability of a pair of particles chosen from different events, which is equal to the product of the single particle probability densities $P(y_1)$ and $P(y_2)$. \footnote{$P_\textmd{diff}(y_1, y_2) = P_1(y_1)P(y_2)$ is only suitable for the discussion under the assumption: for a centrality window, the shapes of mean single-particle probability density distribution $P_1(y)$ of different centralities are similar with each other. In other words, the C$_\textmd{definition}$ method should be used for a centrality window which is not too wide. In our test with the {\sc ampt} model of Au+Au collisions at $\sqrt{s_{NN}} = 200~\mathrm{GeV}$, when the centrality window is not wider than 5\%, it works well.}
$P_\textmd{same}$ and $P_\textmd{diff}$ can be expressed as
\begin{eqnarray}
P_\textmd{same}\left(y_1, y_2\right)\!\! &=& \!\!
\frac{\sum_i \! N_i\left(y_1\right) N_i\left(y_2\right) \!-\! \delta(y_1 \!-\! y_2) N_i\left(y_1\right)}
{\delta y_1 \delta y_2 \sum_i n_{{\textmd{ch}},i}\left(n_{{\textmd{ch}},i}-1\right)},~~~~~~
\label{equ:Psame}\\
P_\textmd{diff}\left(y_1, y_2\right) &=&
\frac{\sum_{i\neq j} N_i\left(y_1\right) N_j\left(y_2\right)}
{\delta y_1 \delta y_2 \sum_{i\neq j} n_{\textmd{ch},i}n_{\textmd{ch},j}},
\label{equ:Pdiff}
\end{eqnarray}
where $i$ and $j$ stand for the event numbers. $\delta y_1$ and $\delta y_2$ are the widths of the rapidity bins around $y_1$ and $y_2$, respectively. $N(y_1)$ and $N(y_2)$ are the numbers of particles falling into $y_1$- and $y_2$-bins. $n_\textmd{ch}$ is the number of charged particles in an event in rapidity range $[-Y, Y]$ (In this paper, $Y$ = 6). By taking Eqs.~(\ref{equ:Psame}) and (\ref{equ:Pdiff}) into Eq.~(\ref{equ:definitionOfC2}), two-particle rapidity correlation function can be calculated with
\begin{equation}
C(y_1,y_2)\!
=\!\!
\Big[\frac{\left< N(y_1)N(y_2) \right>}{\left< N(y_1)\right>\left< N(y_2)\right>}
-\frac{\delta (y_1-y_2)}{\left< N(y_1)\right>}\Big]
\frac{\left< n_\textmd{ch}\right>^2}
         {\left< n_\textmd{ch}\!\left(n_\textmd{ch}\!-\!1\right)\right>},
\label{equ:Cdefinition}
\end{equation}
when the number of events is big enough.
And Eq.~(\ref{equ:Cdefinition}) could be understood as
$C(y_1, y_2) =
\frac{\rho_2\left(y_1, y_2\right)}{\rho_1\left(y_1\right)\rho_1\left(y_2\right)}
\cdot\frac{\left< n_\textmd{ch}\right>^2}
{\left< n_\textmd{ch}\!\left(n_\textmd{ch}\!-\!1\right)\right>}$.

\subsection{C$_\textmd{relative}$ method}

In this part, we try to introduce a relative multiplicity to calculate $C(y_1, y_2)$ without the influence of centrality fluctuation (or centrality span) in a centrality window.
\footnote{The core idea of the relative multiplicity was raised in the research about removing or reducing the influence of centrality fluctuation on forward-backward multiplicity correlation strength \cite{rhhe20162}.}

From the discussion in Sec.~\ref{sec:additionalC}, we consider that Eq.~(\ref{equ:CRho}) is suitable for calculating $C(y_1, y_2)$ at a specific centrality (real centrality) or a centrality interval which is narrow enough. The single- and double-particle densities in rapidity can be calculated as
\begin{equation}
\begin{aligned}
&\rho_1(y_1) =  \left<N(y_1)\right>/\delta y_1,~ \rho_1(y_2) =  \left<N(y_2)\right>/\delta y_2,\\
&\rho_2(y_1, y_2) = \left<N(y_1)N(y_2)\right>/\left(\delta y_1 \delta y_2\right),\\
&\rho_2(y_1, y_1) = \left<N(y_1)\left[N(y_1)-1\right]\right>/\left(\delta y_1\right)^2
\end{aligned}
\end{equation}
where $N(y)$ is the number of particles in the rapidity bin around $y$, and $\delta y$ is the bin width. Therefore, for a specific centrality, 2PC can be calculated with
\begin{equation}
C(y_1, y_2; \varepsilon) = \frac{\left<N(y_1)N(y_2)\right>_\varepsilon}
{\left<N(y_1)\right>_\varepsilon \left<N(y_2)\right>_\varepsilon}
-\frac{\delta(y_1 - y_2)}{\left<N(y_1)\right>_\varepsilon},
\label{equ:Cvarepsilon}
\end{equation}
where $\varepsilon$ stands for the specific centrality, and the averaging over the events is denoted by $\left<\cdots\right>\varepsilon$.

\begin{figure}[!t]
\includegraphics[width=8.6cm]{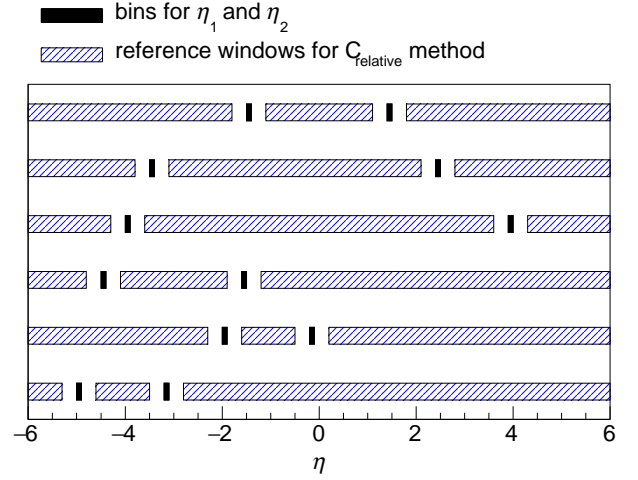}
\caption{\label{fig:illNref}
Illustration of reference windows in the C$_\textmd{relative}$ method.}
\end{figure}

The collision centrality depends on the impact parameter. But in experiments, the impact parameter cannot be determined directly, so in this method, charged particle multiplicity within a rapidity range (which does not overlap the $y_1$ and $y_2$ bins) was used, and called reference multiplicity $N_\textmd{ref}$ \cite{2013profile}. The reference range is illustrated in Fig.~\ref{fig:illNref}.

In this method, reference multiplicity is used not only for centrality cut. For a certain centrality $\varepsilon$, the expectation of reference multiplicity is denoted by $\mu_\textmd{ref}$. Therefore, 2PC at $\varepsilon$ can be expressed as
\begin{equation}
C(y_1, y_2; \varepsilon) = \frac{\left<n^*(y_1) n^*(y_2)\right>_\varepsilon}
{\left<n^*(y_1)\right>_\varepsilon \left<n^*(y_2)\right>_\varepsilon}
-\frac{\delta(y_1 - y_2)}{\mu_\textmd{ref}\left<n^*(y_1)\right>_\varepsilon},
\label{equ:Cvarepsilon2}
\end{equation}
where $n^*(y_1) \equiv N(y_1) / \mu_\textmd{ref}$ and $n^*(y_2) \equiv N(y_2) / \mu_\textmd{ref}$, and are called ideal relative multiplicity.
In a centrality window (Here, it is a reference multiplicity window in fact), if we assume that the shapes of rapidity probability distributions of different centralities are similar with each other, $\left<n^*(y)\right>_\varepsilon = \left<n^*(y)\right>$. Therefore, 2PC of the centrality window can be expressed as (discussed in detail in Appendix~\ref{app:Crelative})
\begin{equation}
C(y_1, y_2) = \frac{\left<n^*(y_1) n^*(y_2)\right>}
{\left<n^*(y_1)\right> \left<n^*(y_2)\right>}
-\frac{\delta(y_1 - y_2)}{\mu_\textmd{ref}\left<n^*(y_1)\right>},
\label{equ:Cvarepsilon3}
\end{equation}
where $\left<\cdots\right>$ without a sub-index stands for the average over the centrality window.
But this equation cannot be used for calculating 2PC directly, because for an event, the ideal relative multiplicities $n^*(y_1)$, $n^*(y_2)$ and reference multiplicity expectation $\mu_\textmd{ref}$ cannot be gotten experimentally.

For a measured multiplicity $N_\textmd{ref}$, we assume that its expectation $\mu_\textmd{ref}$ obeys a Gaussian distribution around $N_\textmd{ref}$ with a half-width $\sqrt{N_\textmd{ref}}$. Reference multiplicity (NO "ideal") is defined as $n(y) = N(y) / N_\textmd{ref}$. For a centrality window (reference multiplicity window), the relationship between the averages of "relative multiplicity" and "ideal relative multiplicity" can be expressed as (see Ref.~\cite{rhhe20162} or Appendix~\ref{app:Crelative}).
\begin{equation}
\begin{aligned}
&\left< n(y_1) \right> = \left< n^*(y_1)\right>,~~
\left< n(y_2) \right> = \left< n^*(y_2)\right>,\\
&\left< n(y_1)n(y_2)\right> = \left< n^*(y_1)n^*(y_1)\right> \big(1 + \frac{1}{\left< N_\textmd{ref}\right>}\big),\\
&\left< n^2(y_1) \right>
= \left< {n^*}^2(y_1)\right> \big(1+\frac{1}{\left< N_\textmd{ref}\right>}\big),~~
\big<\frac{1}{N_\textmd{ref}}\big>
=\big<\frac{1}{\mu_\textmd{ref}}\big>,
\end{aligned}
\label{equ:nAndnStar}
\end{equation}
Taking Eq.~(\ref{equ:nAndnStar}) into Eq.~(\ref{equ:Cvarepsilon3}), 2PC can be expressed as
\begin{equation}
C\left(y_1,y_2\right)
\!=\! \frac{\left< n(y_1) n(y_2)\right>}
             {\big(1 \!+\! \frac{1}{\left< N_\textmd{ref}\right>}\big)
                  \left< n(y_1) \right> \left< n(y_2) \right>}
 - \frac{\delta(y_1\! -\!y_2)}{\left< n(y_1) \right>} \big<\!\frac{1}{N_\textmd{ref}}\!\big>,
\label{equ:Crelative}
\end{equation}
where relative multiplicities $n(y_1) \equiv N(y_1) / N_\textmd{ref}$, $n(y_2) \equiv N(y_2) / N_\textmd{ref}$ and reference multiplicity $N_\textmd{ref}$ can be measured experimentally.

\begin{figure*}
\includegraphics[width=17.2cm]{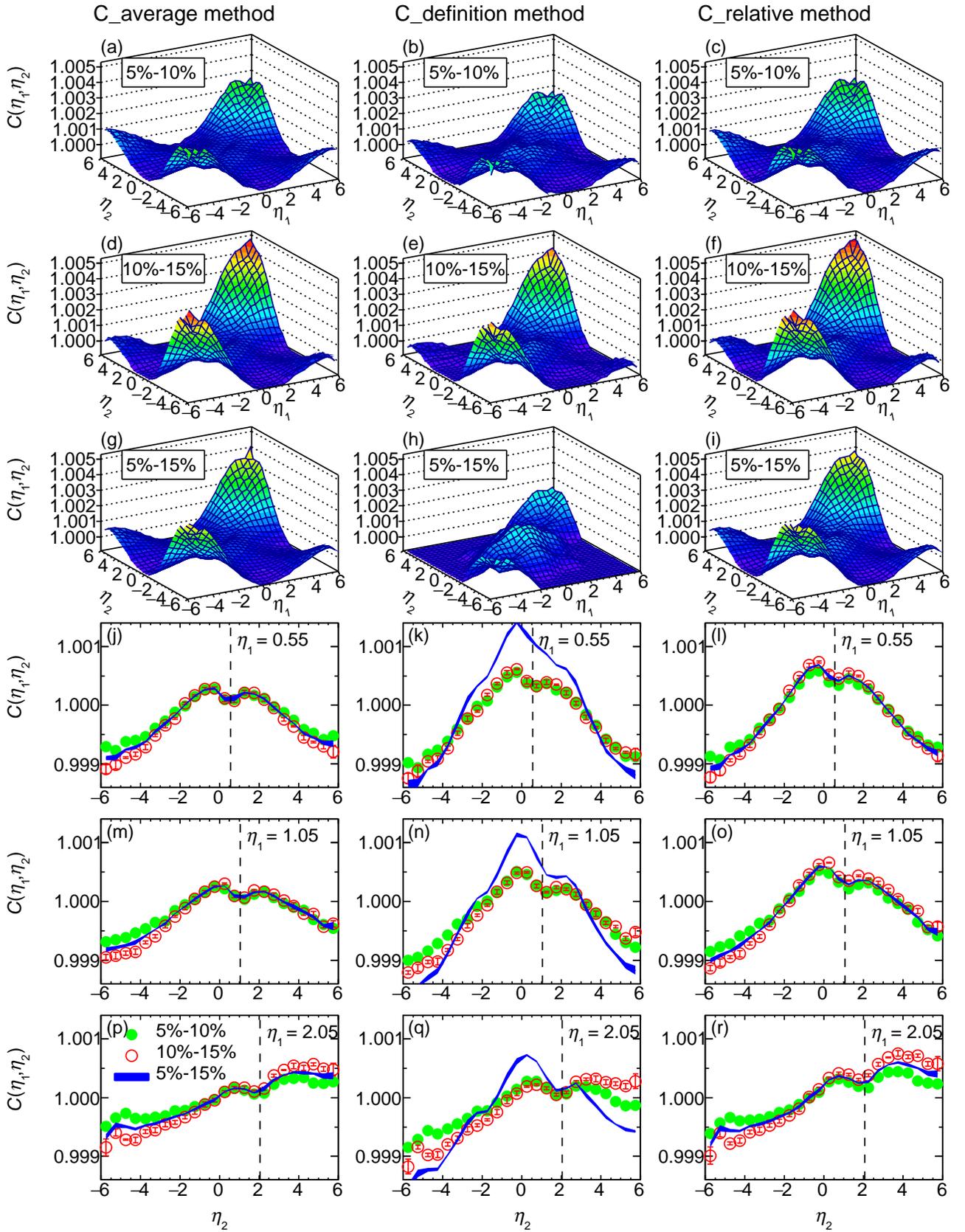}
\caption{\label{fig:resAmpt1}
$C(\eta_1, \eta_2)$ calculated with C$_\textmd{average}$\cite{2016prcc2}, C$_\textmd{definition}$, and C$_\textmd{relative}$ method are drawn in the 1st, 2nd, and 3th columns. In subgraph (a) to (i), $C(\eta_1, \eta_2)$ of 5\%-10\%, 10\%-15\%, and 5\%-15\% are shown. In subgraph (j) to (r), for a specific $\eta_1$, $C(\eta_1, \eta_2)$ of 5\%-10\%, 10\%-15\%, and 5\%-15\% are shown, which can be understood as a sectional view of $C(\eta_1, \eta_2)$, and $\eta_1$ is signed with a dashed line. }
\end{figure*}

\section{testing the methods with {\sc ampt} model and a toy granular model\label{sec:granularModel}}

In Fig.~\ref{fig:resAmpt1}, for the string melting {\sc ampt} model of Au+Au collisions at $\sqrt{s_{NN}} = 200~\mathrm{GeV}$ (partonic scattering section is $3~\mathrm{mb}$), two-particle pseudorapidity correlation function $C(\eta_1, \eta_2)$ of centrality windows 5\%-15\%, 5\%-10\%, and 10\%-15\% are calculated with three methods. For the C$_\textmd{average}$\footnote{This method was raised in Ref.~\cite{2016prcc2} and introduced simply in Sec.~\ref{sec:introduction} of this paper.}, C$_\textmd{definition}$, and C$_\textmd{relative}$ methods, the centrality windows are cut with multiplicity of all particles, number of charged hadrons within pseudorapidity range $[-Y, Y]$, and reference multiplicity $N_\textmd{ref}$, respectively.

It is seen obviously in Fig.~\ref{fig:resAmpt1} that 2PCs are all very week and around 1, and the results with different methods are similar with each other. For the C$_\textmd{average}$ and C$_\textmd{relative}$ methods, 2PC of 5\%-15\% is in the middle of 2PCs of 5\%-10\% and 10\%-15\%, and this phenomenon is more reasonable than the results in Fig.~\ref{fig:ampt}, which are calculated with the ratio of $\frac{\rho_2(\eta_1, \eta_2)}{\rho_1(\eta_1)\rho_1(\eta_2)}$ [Eq.~(\ref{equ:CRho})] directly. Unfortunately, the C$_\textmd{defintion}$ method is suitable for the centrality windows wider than 5\%. By the way, the vertical scale in Fig.~\ref{fig:resAmpt1} is different from Fig.~\ref{fig:ampt}. In other words, for {\sc ampt} model, $C(\eta_1, \eta_2)$ calculated with the C$_\textmd{average}$, C$_\textmd{definition}$, and C$_\textmd{relative}$ methods are weaker than the results in Fig.~\ref{fig:resAmpt1}. In addition, they are closer to 1, and even lower than 1 (negative correlation).

\begin{figure}[!t]
\includegraphics[width=8.6cm]{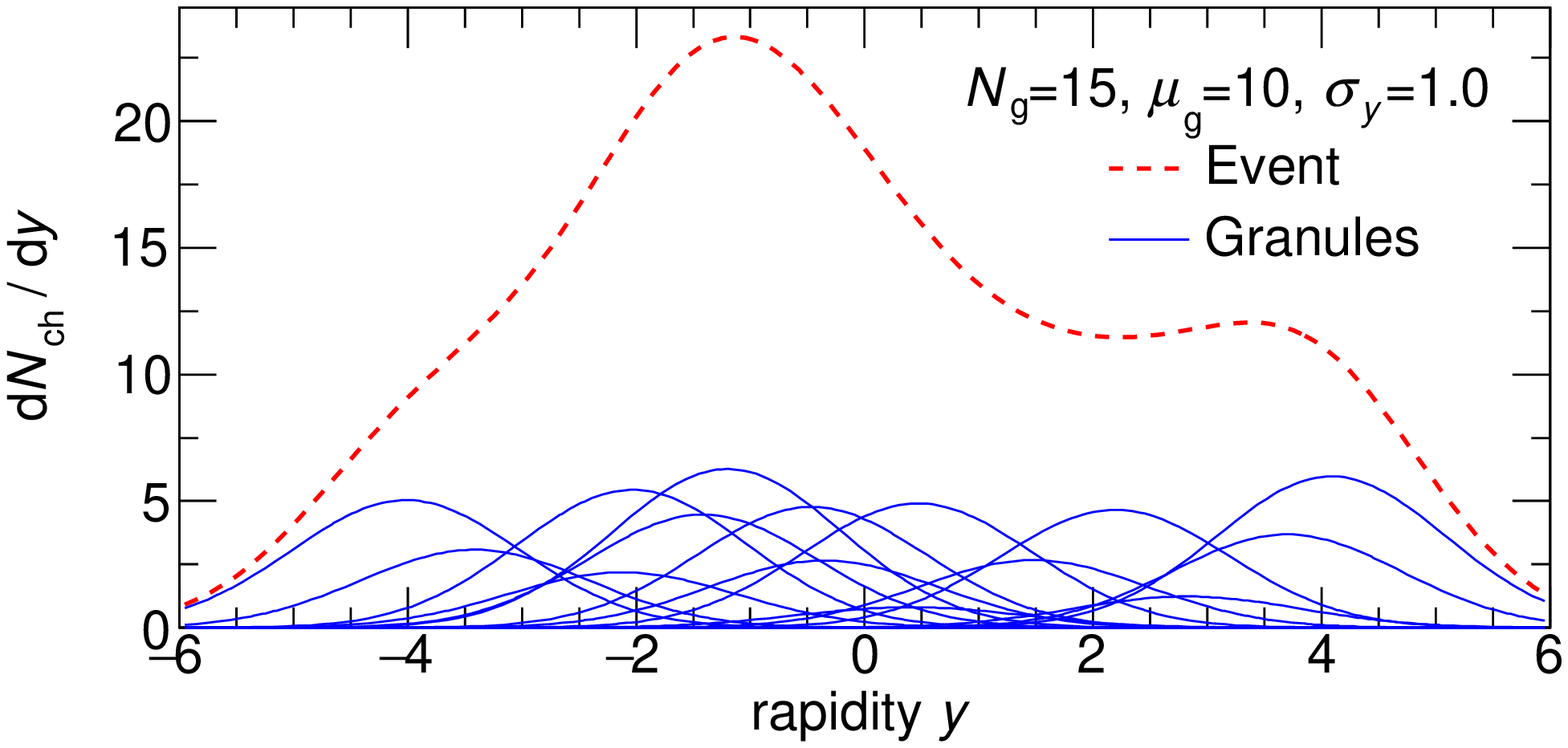}
\caption{\label{fig:illGran} For the granular model, when $N_g$, $\mu_g$, and $\sigma_y$ are set to 15, 10, and 1.0, respectively, the probability density distributions of granules in a random event (blue full curve)  and a total probability density distribution of the event (red dashed curve) are shown.}
\end{figure}

On the other hand, we build a toy granular model to simulate the particle density fluctuation in rapidity. This idea is borrowed from some researches about two-pion correlation \cite{WNZhangGran1, WNZhangGran2, WNZhangGran3, WNZhangGran4} and some similar simulation based on blast-wave model \cite{paper9added1, paper9added2, paper9added3}.  We assume that an event is composed of some random granules, and the particles in a granule share the same collective rapidity. Fig.~\ref{fig:illGran} illustrates the particle density distributions of the granules in a random event. In Fig.~\ref{fig:illGran}, $N_g$ is the number of granules, $\mu_g$ is the expectation of number of particles in a granule, $\sigma_y$ is the half-width of the particle density distribution of a granule. Here, the rapidity of particles in a granule obeys a Gaussian distribution around the collective rapidity, and the collective rapidity distribution is borrowed from mean rapidity distribution of {\sc ampt} model used above, though this assumption is rough.

\begin{figure}[!t]
\includegraphics[width=8.5cm]{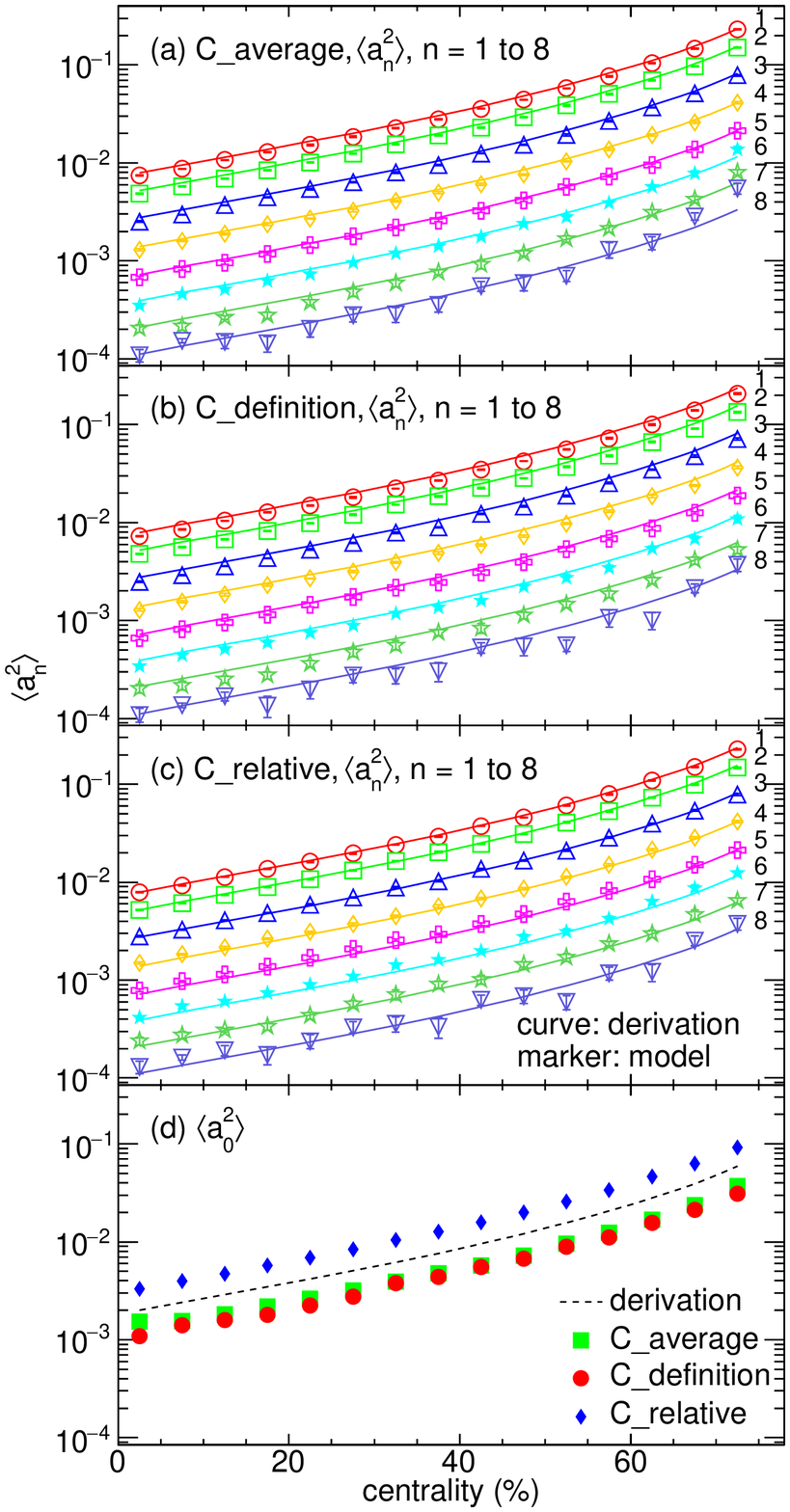}~~
\caption{\label{fig:GranASize}
For the granular model with $\mu_g = 10$ and $\sigma_y = 1$,  $\left<a_n^2 \right>$ ($n$ = 1,2,$\cdots$,8) as a function of centrality are calculated with the C$_\textmd{average}$, C$_\textmd{definition}$, and C$_\textmd{relative}$ methods and drawn as makers in (a), (b), and (c), respectively, and the corresponding deduced results are drawn with curves. In (d), comparisons between $\left<a_0^2\right>$ of derivation and these three methods are shown.}
\end{figure}

More than $2\times 10^5$ random events are made with the granular model (Here, $\mu_g = 10$ and $\sigma_y = 1.0$) by utilizing the distribution of charged hadron multiplicity ($\left|y\right| < 6$) of {\sc ampt} model used above. We deduced the theoretical expression of $\left<a_n^2\right>$ of the granular model in Appendix~\ref{app:granular}, as shown in Eq.~(\ref{equ:gran5}).
For testing the 2PC calculation methods, $\left<a_n^2\right>$ as a function of centrality \footnote{Here, for the C$_\textmd{average}$ and C$_\textmd{definition}$ methods, centrality bins are cut with charged particle multiplicity within $[-Y, Y]$, and for the C$_\textmd{relative}$ method, centrality bins are cut with the reference multiplicity illustrated in Fig.~\ref{fig:illNref}.} is calculated with Eq.~(\ref{equ:aman}) and shown in Fig.~\ref{fig:GranASize}. It's seen that for $n = 1, 2, \cdots, 8$, the $\left<a_n^2\right>$ calculated with C$_\textmd{average}$, C$_\textmd{average}$ and C$_\textmd{average}$ are similar with the results of derivation as shown in subgraph (a), (b), and (c), respectively. But for $n = 0$, the differences between them can be observed as shown in subgraph (d).

In addition, we observe that $\left<a_n^2\right>$ increases with centrality in Fig.~\ref{fig:GranASize}. It can be explained as: when the multiplicity expectation $\mu_g$ of granules is set, for less central events, there are less granules, which may lead to more violent fluctuation of particle density distribution.

\section{summary\label{sec:summary}}

The additional positive correlation is presented in the calculation utilizing the ratio of two-particle density $\rho_2(y_1, y_2)$ to product of single-particle densities $\rho_1(y_1)\rho_1(y_2)$ with {\sc ampt} model of Au+Au collisions at $\sqrt{s_{NN}} = 200~\mathrm{GeV}$. We argue that this phenomenon is due to a mixing of events of different centralities in a centrality window. In other words, this phenomenon may be caused by the centrality span. For removing or deducing the influence of centrality mixing (or centrality span), the C$_\textmd{definition}$ and C$_\textmd{relative}$ methods are raised. In C$_\textmd{definition}$ method, the expression for calculating $C(y_1, y_2)$ is deduced from the normalized ratio of two-particle probability to product of single-particle probability [Eq.~(\ref{equ:definitionOfC})]. In C$_\textmd{relative}$ method, a relative multiplicity is introduced for unifying the events of different centralities, and the bias caused by fluctuation of reference multiplicity is modified under a Gaussian assumption. The methods are tested with {\sc ampt} model and a toy granular model which simulates particle density fluctuation in rapidity. For {\sc ampt} model, 2PC of 5\%-15\% is in the middle of 2PCs of 5\%-10\% and 10\%-15\%, as expected. For the granular model, most of results of theoretical results $\left<a_n^2\right>$ ($n$ = 1, 2, $\dots$, 8) can be reproduced with the calculation methods, though for $\left<a_0^2\right>$, the differences between the results of model and derivation can be seen.

\begin{acknowledgments}
We thank Yan Yang, Hang Yang, Professor Wei-Ning Zhang and Professor Jing-Bo Zhang for the advices and discussions.
\end{acknowledgments}

\appendix

\section{$\left<a_m a_n\right>$ without statistical fluctuation\label{app:aman}}

In this part, we aim at why $\left<a_m a_n\right>$ without statistical fluctuation was written as $\left<a_m^\textmd{obs} a_n^\textmd{obs}\right> -\left< a_m^\textmd{ran} a_n^\textmd{ran} \right>$.

The $y$ distribution of particles in an event can be understood as the sum of statistical fluctuation and the particle density distribution without statistical fluctuation, and the latter can be seen as a collective part, and the former can be seen as a non-collective part. Before the discussion, some functions are defined as follows:

(1) $\rho(y; a_0, a_1, a_2, \cdots)$: the particle density without statistical fluctuation (which can be understood as the "collective" part).

(2) $\rho^\textmd{ran}(y; a_0^\textmd{ran},  a_1^\textmd{ran}, a_2^\textmd{ran}, \cdots)$: the particle density distribution including only statistical fluctuation (which can be understood as "non-collective").

(3) $\rho^\textmd{obs}(y; a_0^\textmd{obs}, a_1^\textmd{obs}, a_2^\textmd{obs}, \cdots)$: the particle density distribution including both event-by-event fluctuation and statistical fluctuation (which can be understood as the sum of "collective" and "non-collective").

(4) $\rho_0(y)$: the mean single-particle density distribution.

For an event,
\begin{eqnarray}
&&\rho^\textmd{obs}(y; a_0^\textmd{obs}, a_1^\textmd{obs}, a_2^\textmd{obs}, \cdots)=\rho_0(y) \Big[1 + \sum_n a_n^\textmd{obs}T_n(y)\Big],\nonumber\\
&&\rho^\textmd{ran}(y; a_0^\textmd{ran},  a_1^\textmd{ran}, a_2^\textmd{ran}, \cdots)=\rho_0(y) \Big[1 + \sum_n a_n^\textmd{ran}T_n(y)\Big],\nonumber\\
&&\rho(y; a_0, a_1, a_2, \cdots)=\rho_0(y) \Big[1 + \sum_n a_nT_n(y)\Big],
\label{equ:AppARho}
\end{eqnarray}
where $T_n(y)$ ( $n$ = 0, 1, 2, $\cdots$) are complete orthogonal basis functions. Because the observed particle density distribution $\rho^\textmd{obs}(y; a_0^\textmd{obs}\!\!\!, a_1^\textmd{obs}\!\!\!, \dots)$ can be understood as the sum of the $\rho(y; a_0, a_1, \dots)$ and pure statistical fluctuation $\rho^\textmd{ran}(y;a_0^\textmd{ran}\!\!\!, a_1^\textmd{ran}\!\!\!, \dots) - \rho_0(y)$. Hence,
\begin{eqnarray}
&&\rho^\textmd{obs}(y; a_0^\textmd{obs}\!\!\!, a_1^\textmd{obs}\!\!\!, \dots)\!\!
=  \rho(y; a_0, a_1, \dots) \nonumber\\
&&~~~~~~~~~~~~~~~~~~~~~~~~
+\!\Big[\rho^\textmd{ran}(y;a_0^\textmd{ran}\!\!\!, a_1^\textmd{ran}\!\!\!, \dots) \!-\!\rho_0(y)\Big],~~~~~\\
&&\Rightarrow
\sum_n a_n^\textmd{obs}T_n(y) = \sum_n a_n^\textmd{ran}T_n(y) + \sum_n a_nT_n(y).~~~
\end{eqnarray}
By utilizing the orthogonality of $T_n(y)$, for an event,
\begin{equation}
a_n^\textmd{obs} = a_n^\textmd{ran} + a_n.
\label{equ:AppARelationshipAnobsAnranAn}
\end{equation}
It is assumed that $a_n^\textmd{ran}$ is independent of $a_m$, no matter whether $m = n$ or $m\neq n$. Therefore,
\begin{equation}
\left< a_m^\textmd{obs}a_n^\textmd{obs}  \right>
\!=\! \left< a_m^\textmd{ran}a_n^\textmd{ran} \right>
\!+\! \left< a_m a_n\right>
\!+\! \left< a_m^\textmd{ran}\right>\left< a_n \right>
\!+\! \left< a_n^\textmd{ran}\right>\left< a_m \right>,
\label{equ:AppA5}
\end{equation}
Besides, from the definition of $\rho^\textmd{obs}$ and $\rho^\textmd{ran}$, it is known
\begin{eqnarray}
&&\left<\rho^\textmd{obs}(y; a_0^\textmd{obs}, a_1^\textmd{obs}, a_2^\textmd{obs}, \cdots)\right>=\rho_0(y),\label{equ:AppA_MeanRhoObs}\\
&&\left<\rho^\textmd{ran}(y; a_0^\textmd{ran},  a_1^\textmd{ran}, a_2^\textmd{ran}, \cdots)\right>=\rho_0(y).\label{equ:AppA_MeanRhoRan}
\end{eqnarray}
Taking the expressions of $\rho^\textmd{obs}$ and $\rho^\textmd{ran}$ in Eq.~(\ref{equ:AppARho}) into Eqs.~(\ref{equ:AppA_MeanRhoObs}) and (\ref{equ:AppA_MeanRhoRan}), respectively,
\begin{equation}
\sum_n \left< a_n^\textmd{obs} \right>T_n(y) = 0, ~~
\sum_n \left< a_n^\textmd{ran} \right>T_n(y) = 0.
\end{equation}
By utilizing the orthogonality of $T_n(y)$ and Eq.~(\ref{equ:AppARelationshipAnobsAnranAn}),
\begin{equation}
\left< a_n^\textmd{obs} \right> = 0,~
\left< a_n^\textmd{ran} \right> = 0, ~
\left< a_n \right> = 0, ~n = 0, 1, 2, \cdots
\label{equ:AppAAnBarIs0}
\end{equation}
Taking Eq.~(\ref{equ:AppAAnBarIs0}) into Eq.~(\ref{equ:AppA5}),
\begin{equation}
\left< a_m^\textmd{obs}a_n^\textmd{obs}  \right>
= \left< a_m^\textmd{ran}a_n^\textmd{ran} \right>
+ \left< a_m a_n\right>.
\label{equ:AppAAmnFinal}
\end{equation}

\section{calculating $\left<a_m a_n\right>$ with two-particle rapidity correlation function\label{app:amanFromC}}

Two-particle rapidity correlation function can be expressed as
\begin{eqnarray}
C(y_1, y_2) \!\!&=&
\frac{\left<\rho^\textmd{obs}\left(y_1; a_1, a_2, \cdots\right)
\rho^\textmd{obs}\left(y_2; a_1, a_2, \cdots\right)\right>}
{\left<\rho^\textmd{obs}\left(y_1; a_1, a_2, \cdots\right)\right>
\left<\rho^\textmd{obs}\left(y_2; a_1, a_2, \cdots\right)\right>}\nonumber\\
&=& 1+\sum_{m, n = 0}^{\infty} \left<a_m^\textmd{obs}a_n^\textmd{obs}\right>T_m(y) T_n(y).
\label{equ:AppB_C}
\end{eqnarray}
For the events including only statistical fluctuation, $C^\textmd{ran}(y_1, y_2) = 1$, so that
\begin{eqnarray}
C^\textmd{ran}(y_1, y_2) &&= 1+\sum_{m, n = 0}^{\infty} \left<a_m^\textmd{ran}a_n^\textmd{ran}\right>T_m(y) T_n(y) = 1, \nonumber\\
\Rightarrow~~~~&& \sum_{m, n = 0}^{\infty} \left<a_m^\textmd{ran}a_n^\textmd{ran}\right>T_m(y) T_n(y) = 0.
\label{equ:AppB_Cran}
\end{eqnarray}
Taking Eq.~(\ref{equ:AppB_Cran}) into Eq.~(\ref{equ:AppB_C}) and utilizing Eq.~(\ref{equ:AppAAmnFinal}),
\begin{equation}
\begin{aligned}
C(y_1, y_2)\!
&=\! 1+\sum_{m, n = 0}^{\infty} \left<a_ma_n\right>T_m(y) T_n(y),\\
&=\! 1\! +\!\!\! \sum_{m, n = 0}^{\infty}\!\!\!\left<a_m a_n\right>
\frac{T_m(y_1)T_n(y_2)\!+\!T_m(y_2)T_n(y_1)}{2}
\end{aligned}
\label{equ:AppB_Cfinal}
\end{equation}
By utilizing the normalization and orthogonality expressed as $\frac{1}{Y^2} \int_{-Y}^{Y} \int_{-Y}^{Y} T_m(y_1)T_n(y_2) dy_1 dy_2 = \delta_{mn}$, integrating both sides of Eq.~(\ref{equ:AppB_Cfinal}), we can get
\begin{equation}
\begin{aligned}
&\frac{1}{Y^2}\int_{-Y}^{Y} \int_{-Y}^{Y} \left[C(y_1, y_2) - 1\right] T_\mu(y_1)T_\nu(y_2) dy_1 dy_2 \\
=&\sum_{m, n = 0}^{\infty} \left<a_m a_n\right> \frac{\delta_{m\mu}\delta_{n\nu}+\delta_{m\nu}\delta_{n\mu}}{2}
=\left<a_\mu a_\nu\right>,
\end{aligned}
\end{equation}
which is equivalent to Eq.~(\ref{equ:aman}).

\section{positive correlation caused by centrality span\label{app:additionalC}}

In this part, we explain where the additional positive correlation is from, when 2PC is calculated with Eq.~(\ref{equ:CRho}). In a centrality window, we assume that the 2PCs of narrow sub-windows are equal to each other. For example, in the centrality window 10\%-20\%, we assume
\begin{equation}
C(y_1, y_2; 10\%) \!=\! C(y_1, y_2; 11\%) \!=\! \cdots \!=\! C(y_1, y_2; 20\%) \!=\! c.
\label{equ:AppCConstant}
\end{equation}
Here, $C(y_1, y_2; \varepsilon)$ stands for the 2PC of a centrality bin around $\varepsilon$, the bin widths are all equal to 1\%, $c$ is a constant.
Hence,
\begin{eqnarray}
&&~~~\begin{aligned}
&\left.\rho_2(y_1, y_2; 10\%)\middle/\big[\rho_1(y_1; 10\%)\rho_1(y_2; 10\%)\big]\right.\\
=&\left.\rho_2(y_1, y_2; 11\%)\middle/\big[\rho_1(y_1; 11\%)\rho_1(y_2; 11\%)\big]\right.\\
=&~~~~~~~~~~~~~\cdots\\
=&\left.\rho_2(y_1, y_2; 20\%)\middle/\big[\rho_1(y_1; 20\%)\rho_1(y_2; 20\%)\big]\right. = c.
\end{aligned}~~~\\
&&\Rightarrow
\frac{\left<\rho_2(y_1, y_2; \varepsilon)\right>}
{\left<\rho_1(y_1; \varepsilon)\rho_1(y_2; \varepsilon)\right>}
=\frac{\left.\sum_{\varepsilon}\rho_2(y_1, y_2; \varepsilon)\right.}
{\left.\sum_{\varepsilon}\rho_1(y_1; \varepsilon)\rho_1(y_2; \varepsilon)\right.}
= c.~~~~~
\end{eqnarray}
Besides, we assume that
\begin{eqnarray}
&&\rho_1(y_1; 10\%) > \rho_1(y_1; 11\%) > \cdots > \rho_1(y_1; 20\%),~~~\\
&&\rho_1(y_2; 10\%) > \rho_1(y_2; 11\%) > \cdots > \rho_1(y_2; 20\%).~~~
\end{eqnarray}
Therefore,
\begin{eqnarray}
&&~~\left<\rho_1(y_1; \varepsilon)\right>\left<\rho_1(y_2; \varepsilon)\right>
< \left<\rho_1(y_1; \varepsilon)\rho_1(y_2; \varepsilon)\right>,\\
\Rightarrow&&~~~
\frac{\left<\rho_2(y_1, y_2; \varepsilon)\right>}
{\left<\rho_1(y_1; \varepsilon)\right>\left<\rho_1(y_2; \varepsilon)\right>}
>\frac{\left<\rho_2(y_1, y_2; \varepsilon)\right>}
{\left<\rho_1(y_1; \varepsilon)\rho_1(y_2; \varepsilon)\right>},~~~\\
\Rightarrow&&~~~~~~~~C(y_1, y_2; 10\%\!\!-\!\!20\%) > c,
\label{equ:AppCAdditionalC}
\end{eqnarray}
where, $\left<\cdots\right>$ stands for the average over all the centrality bins within 10\%-20\%. By comparing Eqs.~(\ref{equ:AppCConstant}) and (\ref{equ:AppCAdditionalC}), we argue that some additional positive correlation may be caused by a wide centrality span.

It is notable that in the derivation above, this proposition is used
\begin{equation}
\left.
\begin{aligned}
a_1 > a_2 > \cdots > a_n > 0\\
b_1 > b_2 > \cdots > b_n > 0
\end{aligned}
\right\}
\Rightarrow \left<a_i\right>\left< b_i\right> < \left< a_i b_i \right>.
\end{equation}
It is proved as follows.
\begin{eqnarray}
&&\left<a_i b_i\right> > \left<a_i\right> \left< b_i\right>\\
\Leftrightarrow &&
    n \sum_{i = 1}^n a_i b_i - \sum_{i = 1}^{n}\sum_{j = 1}^{n} a_i b_j > 0
    \\
\Leftrightarrow &&
    \left\{\!
        \begin{aligned}
        & a_1 (b_1 \!-\! b_1) \!+ \!a_1 (b_1 \!-\! b_2) \!+\! \cdots \!+\! a_1 (b_1 \!-\! b_n)\\
        +& a_2 (b_2 \!-\! b_1) \!+\! a_2 (b_2\! \!- b_2) \!+\! \cdots + a_2 (b_2\! \!- b_n)\\
        +& ~~~~~~~~~~~~~~~\cdots \\
        +& a_n (b_n \!-\! b_1) \!+\! a_n (b_n\! -\! b_2) \!+\! \cdots \!+\! a_n (b_n \!-\! b_n) \!> \!0 \end{aligned}
    \right.
    \\
\Leftrightarrow &&
    \sum_{i = 1}^{n} \sum_{j = i+1}^{n} (a_i - a_j) (b_i - b_j) > 0
    \\
\Leftarrow&&
    \left\{
    \begin{aligned}
        a_1 > a_2 > a_3 > \cdots > a_n > 0\\
        b_1 > b_2 > b_3 > \cdots > b_n > 0
    \end{aligned}
    \right. ~~~~\textmd{proved.}
\end{eqnarray}

\section{detailed derivation of C$_\textmd{relative}$ method \label{app:Crelative}}

In this section, some detailed derivations about the C$_\textmd{relative}$ method are added.
For a certain centrality $\varepsilon$, $C(y_1, y_2; \varepsilon)$ can be expressed as Eq.~(\ref{equ:Cvarepsilon2}), just like
\begin{equation}
\begin{aligned}
C(y_1, y_2; \varepsilon)
&= \frac{\left<n^*(y_1) n^*(y_2)\right>_\varepsilon}
{\left<n^*(y_1)\right>_\varepsilon \left<n^*(y_2)\right>_\varepsilon}
-\frac{\delta(y_1 - y_2)}{\mu_\textmd{ref}\left<n^*(y_1)\right>_\varepsilon}\\
&=\frac{\left<\big[n^*(y_1)-\frac{1}{\mu_\textmd{ref}}\delta(y_1 - y_2)\big] n^*(y_2)\right>_\varepsilon}
{\left<n^*(y_1)\right>_\varepsilon \left<n^*(y_2)\right>_\varepsilon},
\end{aligned}
\label{equ:AppRel1}
\end{equation}
where $n^*(y_1) \equiv N(y_1) / \mu_\textmd{ref}$ and $n^*(y_2) \equiv N(y_2) / \mu_\textmd{ref}$. It is equivalent to
\begin{equation}
\begin{aligned}
&C(y_1, y_2; \textmd{window}) \left<n^*(y_1)\right>_\varepsilon \left<n^*(y_2)\right>_\varepsilon\\
&= \Big<\big[n^*(y_1) - \frac{\delta(y_1 - y_2)}{\mu_\textmd{ref}}\big] n^*(y_2)\Big>_\varepsilon.
\label{equ:AppRel2}
\end{aligned}
\end{equation}
Here, we assume $C(y_1, y_2;\varepsilon )$ of different centralities in the window are equal to each other, and denoted by $C(y_1, y_2; \textmd{window})$. If we assume the shapes of single-particle probability density distribution in rapidity of different centralities are similar with each other,
\begin{equation}
\left<n^*(y)\right>_\varepsilon = \left<n^*(y)\right>
\label{equ:AppRel3}
\end{equation}
where $\left<\cdots\right>$ WITHOUT a sub-index $\varepsilon$ stands for the average over the whole centrality window, and it is equivalent to $\int \cdots f(\varepsilon) d\varepsilon$, where $f(\varepsilon)$ is the centrality probability density distribution in the window.
Averaging both sides of Eq.~(\ref{equ:AppRel2}) over the events in the window and utilizing Eq.~(\ref{equ:AppRel3}),
\begin{equation}
C(y_1, y_2; \textmd{window})
= \frac{\Big<\big[n^*(y_1) - \frac{\delta(y_1 - y_2)}{\mu_\textmd{ref}}\big] n^*(y_2)\Big>}
{\left<n^*(y_1)\right> \left<n^*(y_2)\right>}.
\label{equ:AppRel4}
\end{equation}
It is equivalent to Eq.~(\ref{equ:Cvarepsilon3}).

On the other hand, under the Gaussian assumption, we deduce the relationship between the averages about ideal relative multiplicity $n^*(y)$ and relative multiplicity $n(y)$ as follows.  Here, for a measured $N_\textmd{ref}$, we assume that $\mu_\textmd{ref}$ obeys a Gaussian distribution as
\begin{equation}
\begin{aligned}
G\left(\mu_\textmd{ref};N_\textmd{ref}, \sqrt{N_\textmd{ref}}\right)
\!=\! \frac{1}{\sqrt{2\pi N_\textmd{ref}}}
\exp\Big[ \!-\!\frac{\left(\mu_\textmd{ref}\!-\!N_\textmd{ref}\right)^2}{2N_\textmd{ref}} \Big],
\end{aligned}
\label{equ:AppRel5}
\end{equation}
where $\mu_\textmd{ref}$ and $\sqrt{N_\textmd{ref}}$ can be understood as the truth value and statistical error of a measured $N_\textmd{ref}$, respectively. It is worth to note that the Gaussian approximation is not suitable enough for the most central events such as 0\%-5\%, and it is discussed in detail in Ref.~\cite{rhhe20162}. When the most central windows are avoided, under the Gaussian assumption,
\begin{eqnarray}
\big< \frac{\mu_\textmd{ref}}{N_\textmd{ref}} \big>\!
&=&\!\!\int\!\!\!\int\!\!\!\frac{\mu_\textmd{ref}}{N_\textmd{ref}}
          G\left(\mu_\textmd{ref};N_\textmd{ref}, \sqrt{N_\textmd{ref}}\right)
          \!h\left(N_\textmd{ref}\right)  d N_\textmd{ref}d \mu_\textmd{ref} \nonumber\\
&=&1,\label{equ:AppRel6z}\\
\big< \frac{\mu_\textmd{ref}^2}{N_\textmd{ref}^2} \big>\!
&=&\!\!\int\!\!\!\int\!\!\! \frac{\mu_\textmd{ref}^2}{N_\textmd{ref}^2}
          G\left(\mu_\textmd{ref};N_\textmd{ref}, \sqrt{N_\textmd{ref}}\right)
          h\left(N_\textmd{ref}\right)   d N_\textmd{ref}d \mu_\textmd{ref} \nonumber\\
&=& 1 + \frac{1}{\left< N_\textmd{ref}\right>},
\label{equ:AppRel6}
\end{eqnarray}
where $h\left(N_\textmd{ref}\right)$ is the probability density function of $N_\textmd{ref}$ in a centrality window (in fact, it is a $N_\textmd{ref}$ window). We define an ideal relative reference multiplicity $n_\textmd{ref} \equiv N_\textmd{ref} / \mu_\textmd{ref}$, and Eq.~(\ref{equ:AppRel6}) can be simplified to $\big< \frac{1}{n_\textmd{ref}}\big> = 1$ and $\big< \frac{1}{n_\textmd{ref}^2}\big> = 1 + \frac{1}{\left< N_\textmd{ref}\right>}$.

By utilizing Eqs.~(\ref{equ:AppRel6z}) and (\ref{equ:AppRel6}), the relationship between $\left< n(y_1) \right>$, $\left< n(y_2) \right>$, $\left< n(y_1) n(y_2)\right>$, $\left< n^2(y_1)\right>$, $\left< \frac{1}{N_\textmd{ref}}\right>$ and $\left< n^*(y_1) \right>$, $\left< n^*(y_2) \right>$, $\left< n^*(y_1) n^*(y_2)\right>$, $\left< {n^*}^2(y_1) \right>$, $\left< \frac{1}{n_\textmd{ref}} \right>$ can be expressed as
\begin{widetext}
\begin{eqnarray}
&&\left< n(y_1) \right>
= \left< \frac{N(y_1)}{\mu_\textmd{ref}}
                 \frac{\mu_\textmd{ref}}{N_\textmd{ref}} \right>
= \left< n^*(y_1) \frac{1}{n_\textmd{ref}}\right>
= \left< n^*(y_1)\right> \left< \frac{1}{n_\textmd{ref}}\right>
= \left< n^*(y_1)\right>,\nonumber\\
&&\left< n(y_1) n(y_2)\right>
= \left< \frac{N(y_1) }{\mu_\textmd{ref}}\frac{N(y_2)}{\mu_\textmd{ref}}
                 \frac{\mu_\textmd{ref}^2}{N_\textmd{ref}^2} \right>
= \left< \frac{n^*(y_1)n^*(y_2)}{n_\textmd{ref}^2}\right>
= \left< n^*(y_1)n^*(y_2)\right> \big(1 + \frac{1}{\left< N_\textmd{ref}\right>}\big),\nonumber\\
&&\left< n^2(y_1) \right>
= \left< \frac{N^2(y_1)}{\mu_\textmd{ref}^2}
                 \frac{\mu_\textmd{ref}^2}{N_\textmd{ref}^2} \right>
= \left< {n^*}^2(y_1)\right> \left< \frac{1}{n_\textmd{ref}^2}\right>
= \left< {n^*}^2(y_1)\right> \big(1+\frac{1}{\left< N_\textmd{ref}\right>}\big),\nonumber\\
&&\left<\frac{1}{N_\textmd{ref}}\right>
=\left<\frac{1}{\mu_\textmd{ref}}\frac{\mu_\textmd{ref}}{N_\textmd{ref}}\right>
=\left<\frac{1}{\mu_\textmd{ref}}\right>
    \left<\frac{1}{n_\textmd{ref}}\right>
=\left<\frac{1}{\mu_\textmd{ref}}\right>.
\label{equ:AppRel7}
\end{eqnarray}
Similarly, $\langle n(y_2) \rangle = \langle n^*(y_2) \rangle$.
Here, the correlation between $n^*(y_1)$ (or $n^*(y_2)$) and $n_\textmd{ref}$ is ignored. Taking Eq.~(\ref{equ:AppRel7}) into Eq.~(\ref{equ:AppRel4}) and ignoring the correlation between $n^*(y_1)$ (or $n^*(y_2)$) and $\mu_\textmd{ref}$, the expression of 2PC of the C$_\textmd{relative}$ method [Eq.~(\ref{equ:Crelative})] can be gotten.

\section{theoretical expression of $C(y_1, y_2)$ and $\left<a_m a_n\right>$ of granular model\label{app:granular}}

$C(y_1, y_2)$ and $\left<a_m a_n\right>$ of the granular model are deduced as follows. In this part, $G(x; \mu, \sigma)$ stands for a Gaussian function of $x$ with an expectation $\mu$ and standard deviation $\sigma$. In this toy granular model, we assume:

(1) number of particles $m_g$ in a granule obeys a Gaussian function $G(m_g; \mu_g, \sqrt{\mu_g})$;

(2) the normalized collective rapidity probability distribution function is denoted by $f_g(y_g)$;

(3) rapidity of particle in a granule obeys a Gaussian function around the collective rapidity $y_g$, which is denoted by $G(y; y_g, \sigma_y)$.

For an event with $N_g$ granules, when the granular multiplicity $m_{g_i}$ and the granule rapidity $y_{g_i}$ (where the sub-index $i = 1, 2, \dots, N_g$ is the ordinal number of granules) are known, the single- and double-particle density distributions can be expressed as
\begin{eqnarray}\label{equ:gran1}
\rho_1(y; m_{g_1}, m_{g_2}, \cdots, y_{g_1},y_{g_2},\cdots)&=& \sum_{i = 1}^{N_g} m_{g_i} G\left(y;y_{g_i},\sigma_y\right),\\
\rho_2(y_1, y_2; m_{g_1}, m_{g_2}, \cdots, y_{g_1},y_{g_2},\cdots)
&=& \sum_{i = 1}^{N_g} \sum_{j = 1}^{N_g}  m_{g_i} m_{g_j} G\left(y;y_{g_i},\sigma_y\right) G\left(y;y_{g_j},\sigma_y\right).
\end{eqnarray}
When the granule-multiplicity expectation $\mu_g$ and the granular collective rapidity distribution $f_g(y_{g})$ are both known,
\begin{eqnarray}
&&
\begin{aligned}\label{equ:gran2}
&\rho_1(y)
= \!\!\int\!\!\cdots\!\!\int\!\!
\bigg[\sum_{i = 1}^{N_g} m_{g_i} G\left(y;y_{g_i},\sigma_y\right)\bigg]
\prod_{j=1}^{N_g} f_g(y_{g_j})G\left(m_{g_j};\mu_g,\sqrt{\mu_g}\right)
dy_{g_j} dm_{g_j}
= N_g \mu_g \int G\left(y;y_g,\sigma_y\right)  f_g(y_g) dy_g,
\end{aligned}\\
&&
\begin{aligned}\label{equ:gran2a}
\rho_2(y_1, y_2)
=& \!\!\int\!\!\cdots\!\!\int\!\!\sum_{i = 1}^{N_g}\sum_{j = 1}^{N_g} m_{g_i} m_{g_j} G\left(y_1;y_{g_i},\sigma_y\right)G\left(y_2;y_{g_j},\sigma_y\right)
\prod_{k=1}^{N_g} f_g(y_{g_k})G\left(m_{g_k};\mu_g,\sqrt{\mu_g}\right) dy_{g_k} dm_{g_k}\\
=& N_g\left(N_g - 1\right) \mu_g^2
\int G\left(y_1;y_g,\sigma_y\right)  f_g(y_g) dy_g
\int G\left(y_2;y_g,\sigma_y\right)  f_g(y_g) dy_g\\
&+ N_g \left(\mu_g^2 + \mu_g \right)
\int G\left(y_1;y_g,\sigma_y\right) G\left(y_2;y_g,\sigma_y\right)f_g(y_g)dy_g.
\end{aligned}
\end{eqnarray}
The theoretical expression of 2PC of the granular model is calculated as the ratio of $\rho_2(y_1, y_2)$ to $\rho_1(y_1)\rho_1(y_2)$ (Here, the number of granules $N_g$ is a constant),
\begin{equation}\label{equ:gran4}
\begin{aligned}
C(y_1, y_2)
= \frac{\rho_2(y_1, y_2)}{\rho_1(y_1)\rho_1(y_2)}
=  \frac{1}{N_g}\left(1+ \frac{1}{\mu_g}\right)
     \frac{\int G\left(y_1;y_g,\sigma_y\right)
                       G\left(y_2;y_g,\sigma_y\right)f_g(y_g)dy_g}
              {\int G\left(y_1;y_g,\sigma_y\right)  f_g(y_g) dy_g
                \int G\left(y_2;y_g,\sigma_y\right)  f_g(y_g) dy_g}
 + 1 - \frac{1}{N_g}.
\end{aligned}
\end{equation}
Taking Eq.~(\ref{equ:gran4}) into Eq.~(\ref{equ:aman}), $\left<a_m a_n\right>$ can be expressed as
\begin{equation}\label{equ:gran5}
\left<a_m a_n\right>
=\frac{1+\mu_g}{\langle M \rangle Y^2}
\int_{-Y}^{Y}\int_{-Y}^{Y}
     \frac{\int G\left(y_1;y_g,\sigma_y\right)
                       G\left(y_2;y_g,\sigma_y\right)f_g(y_g)dy_g}
              {\int G\left(y_1;y_g,\sigma_y\right)  f_g(y_g) dy_g
                \int G\left(y_2;y_g,\sigma_y\right)  f_g(y_g) dy_g}
T_m(y_1) T_n(y_2)dy_1dy_2
- \frac{2\delta_{m0}\delta_{n0}}{N_g}.
\end{equation}
The factor $\frac{1}{N_g}\left(1+\frac{1}{\mu_g}\right)$ in Eq.~(\ref{equ:gran4}) can be written as $\frac{1 + \mu_g}{\langle M\rangle}$, just like the ratio in Eq.~(\ref{equ:gran5}). Here $\langle M\rangle = N_g\mu_g$ can be understood as mean multiplicity.

\begin{figure*}[!t]
\includegraphics[width=15.5cm]{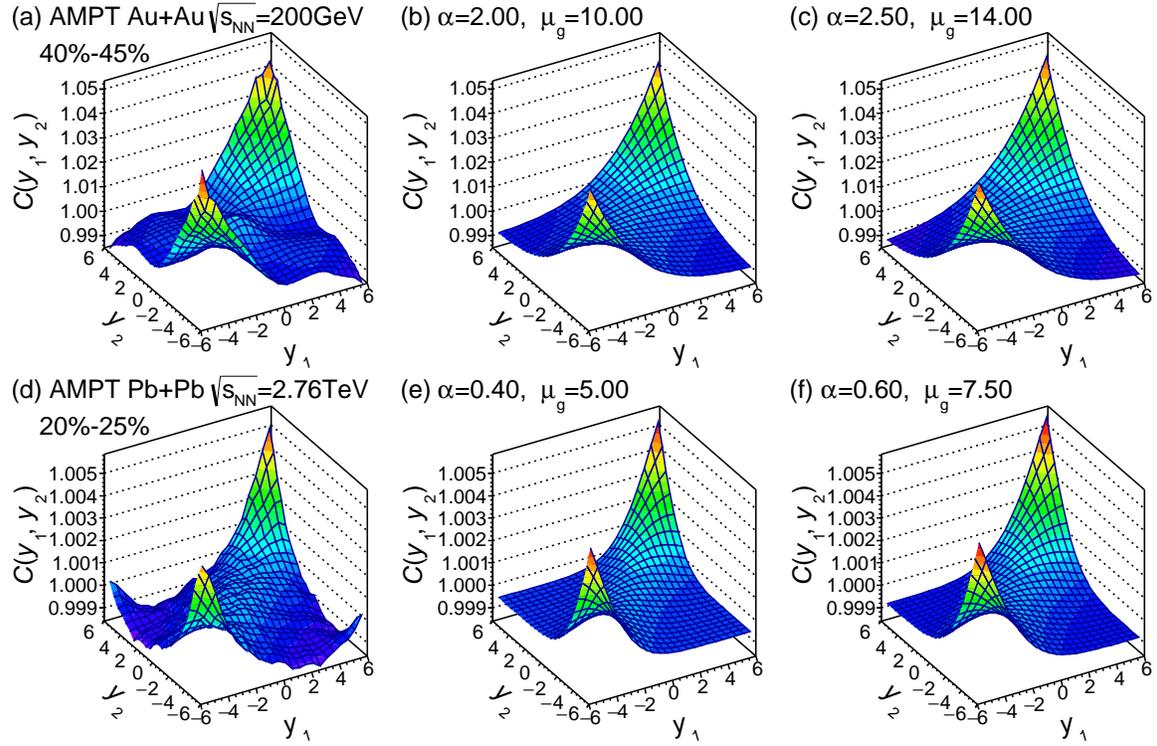}
\caption{\label{fig:GranAuPb}
Comparisons between $C(y_1, y_2)$ of {\sc ampt} model and the deduced results of granular model [Eq.~(\ref{equ:granGau4c})]. For Au+Au collision at $\sqrt{s_{NN}}$ = 200~GeV, 2PC for the centrality window 40\%-45\% are shown in subgraph (a). In subgraph (b) and (c), 2PC are calculated with Eq.~(\ref{equ:granGau4c}) with $\sigma_{\textmd{event}, y}$ and $\langle M\rangle$ which are gotten from corresponding data in (a), and the parameters $\mu_g$ and $\alpha$ are adjusted and signed in the subgraphs. The similar calculations are made for Pb+Pb collisions at $\sqrt{s_{NN}}$ = 2.76~TeV, and the results are shown in (d), (e), and (f).}
\end{figure*}

In addition, for simplifying the expression of $\left<a_m a_n\right>$, we assume that granule-rapidity distribution $f_g(y_y)$ is a Gaussian function with an expectation 0 and a standard deviation $\sigma_{y_g}$, denoted by
\begin{equation}\label{equ:granGau1}
\begin{aligned}
f_g\left(y_g; \sigma_{y_g}\right) = \frac{1}{\sqrt{2\pi}\sigma_{y_g}}\exp{\left(-\frac{y_g^2}{2\sigma_{y_g}^2}\right)}.
\end{aligned}
\end{equation}
For simplifying the following expressions, we define $\alpha \equiv \sigma_y^2 / \sigma_{y_g}^2$, which can be understood as a relative width of granule in rapidity. Taking Eq.~(\ref{equ:granGau1}) into Eqs.~(\ref{equ:gran4}) and (\ref{equ:gran5}), $C(y_1, y_2)$ and $\left<a_m a_n\right>$ can be expressed as
\begin{eqnarray}
\label{equ:granGau2c}
&&C(y_1,y_2)
= \frac{1}{\langle M \rangle}
    \left\{\frac{(1+\mu_g)(\alpha+1)}{\sqrt{\left(\alpha+2\right)\alpha}}
                \exp\left[-\frac{y_1^2+y_2^2-2(\alpha+1)y_1 y_2}{2\sigma_{y_g}^2 \alpha(\alpha+1)(\alpha+2)}\right] - \mu_g\right\}
 + 1,\\
\label{equ:granGau2}
&&\left<a_m a_n\right>
= \frac{1+\mu_g}{\langle M \rangle} \frac{1}{Y^2}
\int \frac{\alpha+1} {\sqrt{(\alpha+2)\alpha}}
        \exp\left[{-\frac{y_1^2+y_2^2-2(\alpha+1)y_1 y_2}{2\sigma_{y_g}^2 \alpha(\alpha+1)(\alpha+2)}}\right]
 T_n(y_1) T_n(y_2)  dy_1 dy_2 - \frac{2\mu_g\delta_{m0}\delta_{n0}}{\left<M\right>}.
\end{eqnarray}
Under the Gaussian assumption of $f_g(y_g)$, the rapidity distribution in events can be expressed as
\begin{equation}\label{equ:granGau3}
\frac{dN_\textmd{ch}}{dy}
\propto\int G(y; y_g, \sigma_y) f_g(y_g;\sigma_{y_g}) dy_g
=\frac{1}{\sqrt{2\pi}\sqrt{\alpha+1}\sigma_{y_g}}
  \exp\left[ -\frac{y^2}{2(\alpha+1)\sigma_{y_g}^2}\right],
\end{equation}
which can be understood as a Gaussian distribution with a half-width $\sigma_{\textmd{event}, y}= \sqrt{\alpha+1}\sigma_{y_g}$. Taking $\sigma_{\textmd{event}, y}$ into Eqs.~(\ref{equ:granGau2c}) and (\ref{equ:granGau2}), $C(y_1, y_2)$ and $\left<a_m a_n\right>$ are expressed respectively as
\begin{eqnarray}
\label{equ:granGau4c}
&&C(y_1,y_2)
= \frac{1}{\langle M \rangle}
    \left\{\frac{(1+\mu_g)(\alpha+1)}{\sqrt{\left(\alpha+2\right)\alpha}}
                \exp\left[-\frac{y_1^2+y_2^2-2(\alpha+1)y_1 y_2}{2\sigma_{\textmd{event}, y}^2 \alpha(\alpha+2)}\right] - \mu_g\right\}
 + 1,\\
\label{equ:granGau4}
&&\left<a_m a_n\right>
= \frac{1+\mu_g}{\langle M \rangle} \frac{1}{Y^2}
\int \frac{\alpha+1} {\sqrt{(\alpha+2)\alpha}}
        \exp\left[{-\frac{y_1^2+y_2^2-2(\alpha+1)y_1 y_2}{2\sigma_{\textmd{event}, y}^2 \alpha(\alpha+2)}}\right]
 T_n(y_1) T_n(y_2)  dy_1 dy_2 - \frac{2\mu_g\delta_{m0}\delta_{n0}}{\left<M\right>},
\end{eqnarray}
where the $\sigma_{\textmd{event}, y}$ and $\left<M\right>$ can be measured as the standard deviation of rapidity and mean charged particle multiplicity, respectively. For {\sc ampt} model of Au+Au collisions at $\sqrt{s_{NN}}$ = 200~GeV (centrality 40\%-45\%) and Pb+Pb collisions at $\sqrt{s_{NN}}$ = 2.76~TeV (centrality 20\%-25\%), we try to reproduce the $C(y_1, y_2)$ by adjusting the parameters $\alpha$ and $\mu_g$, and the results are shown in Fig.~\ref{fig:GranAuPb}.
\end{widetext}


\begin{thebibliography}{34}%
\makeatletter
\providecommand \@ifxundefined [1]{%
 \@ifx{#1\undefined}
}%
\providecommand \@ifnum [1]{%
 \ifnum #1\expandafter \@firstoftwo
 \else \expandafter \@secondoftwo
 \fi
}%
\providecommand \@ifx [1]{%
 \ifx #1\expandafter \@firstoftwo
 \else \expandafter \@secondoftwo
 \fi
}%
\providecommand \natexlab [1]{#1}%
\providecommand \enquote  [1]{``#1''}%
\providecommand \bibnamefont  [1]{#1}%
\providecommand \bibfnamefont [1]{#1}%
\providecommand \citenamefont [1]{#1}%
\providecommand \href@noop [0]{\@secondoftwo}%
\providecommand \href [0]{\begingroup \@sanitize@url \@href}%
\providecommand \@href[1]{\@@startlink{#1}\@@href}%
\providecommand \@@href[1]{\endgroup#1\@@endlink}%
\providecommand \@sanitize@url [0]{\catcode `\\12\catcode `\$12\catcode
  `\&12\catcode `\#12\catcode `\^12\catcode `\_12\catcode `\%12\relax}%
\providecommand \@@startlink[1]{}%
\providecommand \@@endlink[0]{}%
\providecommand \url  [0]{\begingroup\@sanitize@url \@url }%
\providecommand \@url [1]{\endgroup\@href {#1}{\urlprefix }}%
\providecommand \urlprefix  [0]{URL }%
\providecommand \Eprint [0]{\href }%
\providecommand \doibase [0]{http://dx.doi.org/}%
\providecommand \selectlanguage [0]{\@gobble}%
\providecommand \bibinfo  [0]{\@secondoftwo}%
\providecommand \bibfield  [0]{\@secondoftwo}%
\providecommand \translation [1]{[#1]}%
\providecommand \BibitemOpen [0]{}%
\providecommand \bibitemStop [0]{}%
\providecommand \bibitemNoStop [0]{.\EOS\space}%
\providecommand \EOS [0]{\spacefactor3000\relax}%
\providecommand \BibitemShut  [1]{\csname bibitem#1\endcsname}%
\let\auto@bib@innerbib\@empty
\bibitem [{\citenamefont {Bzdak}\ and\ \citenamefont
  {Teaney}(2013)}]{2013prcc2}%
  \BibitemOpen
  \bibfield  {author} {\bibinfo {author} {\bibfnamefont {A.}~\bibnamefont
  {Bzdak}}\ and\ \bibinfo {author} {\bibfnamefont {D.}~\bibnamefont {Teaney}},\
  }\href {http://link.aps.org/doi/10.1103/PhysRevC.87.024906} {\bibfield
  {journal} {\bibinfo  {journal} {Phys. Rev. C}\ }\textbf {\bibinfo {volume}
  {87}},\ \bibinfo {pages} {024906} (\bibinfo {year} {2013})}\BibitemShut
  {NoStop}%
\bibitem [{\citenamefont {Jia}\ \emph {et~al.}(2016{\natexlab{a}})\citenamefont
  {Jia}, \citenamefont {Radhakrishnan},\ and\ \citenamefont
  {Zhou}}]{2016prcc2}%
  \BibitemOpen
  \bibfield  {author} {\bibinfo {author} {\bibfnamefont {J.}~\bibnamefont
  {Jia}}, \bibinfo {author} {\bibfnamefont {S.}~\bibnamefont {Radhakrishnan}},
  \ and\ \bibinfo {author} {\bibfnamefont {M.}~\bibnamefont {Zhou}},\ }\href
  {http://link.aps.org/doi/10.1103/PhysRevC.93.044905} {\bibfield  {journal}
  {\bibinfo  {journal} {Phys. Rev. C}\ }\textbf {\bibinfo {volume} {93}},\
  \bibinfo {pages} {044905} (\bibinfo {year} {2016}{\natexlab{a}})}\BibitemShut
  {NoStop}%
\bibitem [{\citenamefont {Jia}\ \emph {et~al.}(2016{\natexlab{b}})\citenamefont
  {Jia}, \citenamefont {Radhakrishnan}, \citenamefont {Zhou},\ and\
  \citenamefont {Huo}}]{2016prcc2short}%
  \BibitemOpen
  \bibfield  {author} {\bibinfo {author} {\bibfnamefont {J.}~\bibnamefont
  {Jia}}, \bibinfo {author} {\bibfnamefont {S.}~\bibnamefont {Radhakrishnan}},
  \bibinfo {author} {\bibfnamefont {M.}~\bibnamefont {Zhou}}, \ and\ \bibinfo
  {author} {\bibfnamefont {P.}~\bibnamefont {Huo}},\ }\href {\doibase
  http://dx.doi.org/10.1016/j.nuclphysa.2016.02.069} {\bibfield  {journal}
  {\bibinfo  {journal} {Nucl. Phys. A}\ }\textbf {\bibinfo {volume} {956}},\
  \bibinfo {pages} {401} (\bibinfo {year} {2016}{\natexlab{b}})}\BibitemShut
  {NoStop}%
\bibitem [{\citenamefont {Bozek}\ \emph {et~al.}(2011)\citenamefont {Bozek},
  \citenamefont {Broniowski},\ and\ \citenamefont {Moreira}}]{2016cite12}%
  \BibitemOpen
  \bibfield  {author} {\bibinfo {author} {\bibfnamefont {P.}~\bibnamefont
  {Bozek}}, \bibinfo {author} {\bibfnamefont {W.}~\bibnamefont {Broniowski}}, \
  and\ \bibinfo {author} {\bibfnamefont {J.}~\bibnamefont {Moreira}},\ }\href
  {http://link.aps.org/doi/10.1103/PhysRevC.83.034911} {\bibfield  {journal}
  {\bibinfo  {journal} {Phys. Rev. C}\ }\textbf {\bibinfo {volume} {83}},\
  \bibinfo {pages} {034911} (\bibinfo {year} {2011})}\BibitemShut {NoStop}%
\bibitem [{\citenamefont {Jia}\ and\ \citenamefont {Huo}(2014)}]{2016cite14}%
  \BibitemOpen
  \bibfield  {author} {\bibinfo {author} {\bibfnamefont {J.}~\bibnamefont
  {Jia}}\ and\ \bibinfo {author} {\bibfnamefont {P.}~\bibnamefont {Huo}},\
  }\href {http://link.aps.org/doi/10.1103/PhysRevC.90.034915} {\bibfield
  {journal} {\bibinfo  {journal} {Phys. Rev. C}\ }\textbf {\bibinfo {volume}
  {90}},\ \bibinfo {pages} {034915} (\bibinfo {year} {2014})}\BibitemShut
  {NoStop}%
\bibitem [{\citenamefont {Bhalerao}\ \emph {et~al.}(2015)\citenamefont
  {Bhalerao}, \citenamefont {Ollitrault}, \citenamefont {Pal},\ and\
  \citenamefont {Teaney}}]{2016cite15}%
  \BibitemOpen
  \bibfield  {author} {\bibinfo {author} {\bibfnamefont {R.~S.}\ \bibnamefont
  {Bhalerao}}, \bibinfo {author} {\bibfnamefont {J.-Y.}\ \bibnamefont
  {Ollitrault}}, \bibinfo {author} {\bibfnamefont {S.}~\bibnamefont {Pal}}, \
  and\ \bibinfo {author} {\bibfnamefont {D.}~\bibnamefont {Teaney}},\ }\href
  {http://link.aps.org/doi/10.1103/PhysRevLett.114.152301} {\bibfield
  {journal} {\bibinfo  {journal} {Phys. Rev. Lett.}\ }\textbf {\bibinfo
  {volume} {114}},\ \bibinfo {pages} {152301} (\bibinfo {year}
  {2015})}\BibitemShut {NoStop}%
\bibitem [{\citenamefont {Pang}\ \emph {et~al.}(2015)\citenamefont {Pang},
  \citenamefont {Qin}, \citenamefont {Roy}, \citenamefont {Wang},\ and\
  \citenamefont {Ma}}]{2016cite18}%
  \BibitemOpen
  \bibfield  {author} {\bibinfo {author} {\bibfnamefont {L.-G.}\ \bibnamefont
  {Pang}}, \bibinfo {author} {\bibfnamefont {G.-Y.}\ \bibnamefont {Qin}},
  \bibinfo {author} {\bibfnamefont {V.}~\bibnamefont {Roy}}, \bibinfo {author}
  {\bibfnamefont {X.-N.}\ \bibnamefont {Wang}}, \ and\ \bibinfo {author}
  {\bibfnamefont {G.-L.}\ \bibnamefont {Ma}},\ }\href
  {http://link.aps.org/doi/10.1103/PhysRevC.91.044904} {\bibfield  {journal}
  {\bibinfo  {journal} {Phys. Rev. C}\ }\textbf {\bibinfo {volume} {91}},\
  \bibinfo {pages} {044904} (\bibinfo {year} {2015})}\BibitemShut {NoStop}%
\bibitem [{\citenamefont {Khachatryan}\ \emph {et~al.}(2015)\citenamefont
  {Khachatryan} \emph {et~al.}}]{2016cite19}%
  \BibitemOpen
  \bibfield  {author} {\bibinfo {author} {\bibfnamefont {V.}~\bibnamefont
  {Khachatryan}} \emph {et~al.} (\bibinfo {collaboration} {CMS
  Collaboration}),\ }\href {http://link.aps.org/doi/10.1103/PhysRevC.92.034911}
  {\bibfield  {journal} {\bibinfo  {journal} {Phys. Rev. C}\ }\textbf {\bibinfo
  {volume} {92}},\ \bibinfo {pages} {034911} (\bibinfo {year}
  {2015})}\BibitemShut {NoStop}%
\bibitem [{\citenamefont {Bialas}\ and\ \citenamefont
  {Zalewski}(2010)}]{asymmetry1}%
  \BibitemOpen
  \bibfield  {author} {\bibinfo {author} {\bibfnamefont {A.}~\bibnamefont
  {Bialas}}\ and\ \bibinfo {author} {\bibfnamefont {K.}~\bibnamefont
  {Zalewski}},\ }\href {http://link.aps.org/doi/10.1103/PhysRevC.82.034911}
  {\bibfield  {journal} {\bibinfo  {journal} {Phys. Rev. C}\ }\textbf {\bibinfo
  {volume} {82}},\ \bibinfo {pages} {034911} (\bibinfo {year}
  {2010})}\BibitemShut {NoStop}%
\bibitem [{\citenamefont {Radhakrishnan}(2016)}]{thesis-c-expPbPb}%
  \BibitemOpen
  \bibfield  {author} {\bibinfo {author} {\bibfnamefont {S.}~\bibnamefont
  {Radhakrishnan}} (\bibinfo {collaboration} {ATLAS}),\ }in\ \href {\doibase
  10.1016/j.nuclphysbps.2016.05.024} {\emph {\bibinfo {booktitle}
  {{Proceedings, 7th International Conference on Hard and Electromagnetic
  Probes of High-Energy Nuclear Collisions (Hard Probes 2015): Montréal,
  Québec, Canada, June 29-July 3, 2015}}}}\ (\bibinfo {year} {2016})\ \Eprint
  {http://arxiv.org/abs/1511.00361} {arXiv:1511.00361 [nucl-ex]} \BibitemShut
  {NoStop}%
\bibitem [{\citenamefont {Gyulassy}\ \emph {et~al.}(1979)\citenamefont
  {Gyulassy}, \citenamefont {Kauffmann},\ and\ \citenamefont
  {Wilson}}]{hbtprc1979}%
  \BibitemOpen
  \bibfield  {author} {\bibinfo {author} {\bibfnamefont {M.}~\bibnamefont
  {Gyulassy}}, \bibinfo {author} {\bibfnamefont {S.~K.}\ \bibnamefont
  {Kauffmann}}, \ and\ \bibinfo {author} {\bibfnamefont {L.~W.}\ \bibnamefont
  {Wilson}},\ }\href {http://link.aps.org/doi/10.1103/PhysRevC.20.2267}
  {\bibfield  {journal} {\bibinfo  {journal} {Phys. Rev. C}\ }\textbf {\bibinfo
  {volume} {20}},\ \bibinfo {pages} {2267} (\bibinfo {year}
  {1979})}\BibitemShut {NoStop}%
\bibitem [{\citenamefont {Ravan}\ \emph {et~al.}(2014)\citenamefont {Ravan},
  \citenamefont {Pujahari}, \citenamefont {Prasad},\ and\ \citenamefont
  {Pruneau}}]{added4}%
  \BibitemOpen
  \bibfield  {author} {\bibinfo {author} {\bibfnamefont {S.}~\bibnamefont
  {Ravan}}, \bibinfo {author} {\bibfnamefont {P.}~\bibnamefont {Pujahari}},
  \bibinfo {author} {\bibfnamefont {S.}~\bibnamefont {Prasad}}, \ and\ \bibinfo
  {author} {\bibfnamefont {C.~A.}\ \bibnamefont {Pruneau}},\ }\href
  {http://link.aps.org/doi/10.1103/PhysRevC.89.024906} {\bibfield  {journal}
  {\bibinfo  {journal} {Phys. Rev. C}\ }\textbf {\bibinfo {volume} {89}},\
  \bibinfo {pages} {024906} (\bibinfo {year} {2014})}\BibitemShut {NoStop}%
\bibitem [{\citenamefont {Vechernin}(2015)}]{CorrFuncVV2015}%
  \BibitemOpen
  \bibfield  {author} {\bibinfo {author} {\bibfnamefont {V.}~\bibnamefont
  {Vechernin}},\ }\href {\doibase
  http://dx.doi.org/10.1016/j.nuclphysa.2015.03.009} {\bibfield  {journal}
  {\bibinfo  {journal} {Nucl. Phys. A}\ }\textbf {\bibinfo {volume} {939}},\
  \bibinfo {pages} {21} (\bibinfo {year} {2015})}\BibitemShut {NoStop}%
\bibitem [{\citenamefont {Monnai}\ and\ \citenamefont
  {Schenke}(2016)}]{CorrFuncPLB2016}%
  \BibitemOpen
  \bibfield  {author} {\bibinfo {author} {\bibfnamefont {A.}~\bibnamefont
  {Monnai}}\ and\ \bibinfo {author} {\bibfnamefont {B.}~\bibnamefont
  {Schenke}},\ }\href {\doibase
  http://dx.doi.org/10.1016/j.physletb.2015.11.063} {\bibfield  {journal}
  {\bibinfo  {journal} {Phys. Lett. B}\ }\textbf {\bibinfo {volume} {752}},\
  \bibinfo {pages} {317} (\bibinfo {year} {2016})}\BibitemShut {NoStop}%
\bibitem [{\citenamefont {Pruneau}\ \emph {et~al.}(2002)\citenamefont
  {Pruneau}, \citenamefont {Gavin},\ and\ \citenamefont
  {Voloshin}}]{CorrFuncPRC2002}%
  \BibitemOpen
  \bibfield  {author} {\bibinfo {author} {\bibfnamefont {C.}~\bibnamefont
  {Pruneau}}, \bibinfo {author} {\bibfnamefont {S.}~\bibnamefont {Gavin}}, \
  and\ \bibinfo {author} {\bibfnamefont {S.}~\bibnamefont {Voloshin}},\ }\href
  {http://link.aps.org/doi/10.1103/PhysRevC.66.044904} {\bibfield  {journal}
  {\bibinfo  {journal} {Phys. Rev. C}\ }\textbf {\bibinfo {volume} {66}},\
  \bibinfo {pages} {044904} (\bibinfo {year} {2002})}\BibitemShut {NoStop}%
\bibitem [{\citenamefont {Khachatryan}\ \emph {et~al.}(2010)\citenamefont
  {Khachatryan} \emph {et~al.}}]{CorrFuncJHEP2010}%
  \BibitemOpen
  \bibfield  {author} {\bibinfo {author} {\bibfnamefont {V.}~\bibnamefont
  {Khachatryan}} \emph {et~al.} (\bibinfo {collaboration} {CMS}),\ }\href
  {\doibase 10.1007/JHEP09(2010)091} {\bibfield  {journal} {\bibinfo  {journal}
  {J. High Energy Phys.}\ }\textbf {\bibinfo {volume} {09}},\ \bibinfo {pages}
  {091} (\bibinfo {year} {2010})},\ \Eprint {http://arxiv.org/abs/1009.4122}
  {arXiv:1009.4122 [hep-ex]} \BibitemShut {NoStop}%
\bibitem [{\citenamefont {Vechernin}(2013)}]{vv2014}%
  \BibitemOpen
  \bibfield  {author} {\bibinfo {author} {\bibfnamefont {V.}~\bibnamefont
  {Vechernin}},\ }\href@noop {} {\bibfield  {journal} {\bibinfo  {journal}
  {arXiv}\ }\textbf {\bibinfo {volume} {1305.0857}} (\bibinfo {year}
  {2013})}\BibitemShut {NoStop}%
\bibitem [{\citenamefont {Lin}\ \emph {et~al.}(2005)\citenamefont {Lin},
  \citenamefont {Ko}, \citenamefont {Li}, \citenamefont {Zhang},\ and\
  \citenamefont {Pal}}]{ampt}%
  \BibitemOpen
  \bibfield  {author} {\bibinfo {author} {\bibfnamefont {Z.-W.}\ \bibnamefont
  {Lin}}, \bibinfo {author} {\bibfnamefont {C.~M.}\ \bibnamefont {Ko}},
  \bibinfo {author} {\bibfnamefont {B.-A.}\ \bibnamefont {Li}}, \bibinfo
  {author} {\bibfnamefont {B.}~\bibnamefont {Zhang}}, \ and\ \bibinfo {author}
  {\bibfnamefont {S.}~\bibnamefont {Pal}},\ }\href
  {http://link.aps.org/doi/10.1103/PhysRevC.72.064901} {\bibfield  {journal}
  {\bibinfo  {journal} {Phys. Rev. C}\ }\textbf {\bibinfo {volume} {72}},\
  \bibinfo {pages} {064901} (\bibinfo {year} {2005})}\BibitemShut {NoStop}%
\bibitem [{\citenamefont {He}\ \emph {et~al.}(2016{\natexlab{a}})\citenamefont
  {He}, \citenamefont {Qian},\ and\ \citenamefont {Huo}}]{rhhe2016}%
  \BibitemOpen
  \bibfield  {author} {\bibinfo {author} {\bibfnamefont {R.}~\bibnamefont
  {He}}, \bibinfo {author} {\bibfnamefont {J.}~\bibnamefont {Qian}}, \ and\
  \bibinfo {author} {\bibfnamefont {L.}~\bibnamefont {Huo}},\ }\href
  {http://dx.doi.org/10.1103/PhysRevC.93.044918} {\bibfield  {journal}
  {\bibinfo  {journal} {Phys. Rev. C}\ }\textbf {\bibinfo {volume} {93}},\
  \bibinfo {pages} {044918} (\bibinfo {year} {2016}{\natexlab{a}})}\BibitemShut
  {NoStop}%
\bibitem [{\citenamefont {Yan}\ \emph {et~al.}(2009)\citenamefont {Yan},
  \citenamefont {Zhou}, \citenamefont {Dong}, \citenamefont {Li}, \citenamefont
  {Ma},\ and\ \citenamefont {Sa}}]{2009bOfCentralityAndAAModel}%
  \BibitemOpen
  \bibfield  {author} {\bibinfo {author} {\bibfnamefont {Y.-L.}\ \bibnamefont
  {Yan}}, \bibinfo {author} {\bibfnamefont {D.-M.}\ \bibnamefont {Zhou}},
  \bibinfo {author} {\bibfnamefont {B.-G.}\ \bibnamefont {Dong}}, \bibinfo
  {author} {\bibfnamefont {X.-M.}\ \bibnamefont {Li}}, \bibinfo {author}
  {\bibfnamefont {H.-L.}\ \bibnamefont {Ma}}, \ and\ \bibinfo {author}
  {\bibfnamefont {B.-H.}\ \bibnamefont {Sa}},\ }\href
  {http://adsabs.harvard.edu/abs/2009PhRvC..79e4902Y} {\bibfield  {journal}
  {\bibinfo  {journal} {Phys. Rev. C}\ }\textbf {\bibinfo {volume} {79}},\
  \bibinfo {pages} {054902} (\bibinfo {year} {2009})}\BibitemShut {NoStop}%
\bibitem [{\citenamefont {Olszewski}\ and\ \citenamefont
  {Broniowski}(2016)}]{FB2017}%
  \BibitemOpen
  \bibfield  {author} {\bibinfo {author} {\bibfnamefont {A.}~\bibnamefont
  {Olszewski}}\ and\ \bibinfo {author} {\bibfnamefont {W.}~\bibnamefont
  {Broniowski}},\ }\href@noop {} {\bibfield  {journal} {\bibinfo  {journal}
  {arxiv:}\ }\textbf {\bibinfo {volume} {1701}},\ \bibinfo {pages} {00099}
  (\bibinfo {year} {2016})}\BibitemShut {NoStop}%
\bibitem [{\citenamefont {Fu}(2008)}]{2008bOfEmitRandomly}%
  \BibitemOpen
  \bibfield  {author} {\bibinfo {author} {\bibfnamefont {J.}~\bibnamefont
  {Fu}},\ }\href {http://link.aps.org/doi/10.1103/PhysRevC.77.027902}
  {\bibfield  {journal} {\bibinfo  {journal} {Phys. Rev. C}\ }\textbf {\bibinfo
  {volume} {77}},\ \bibinfo {pages} {027902} (\bibinfo {year}
  {2008})}\BibitemShut {NoStop}%
\bibitem [{\citenamefont {Wiedemann}\ and\ \citenamefont {Heinz}(1999)}]{hbt1}%
  \BibitemOpen
  \bibfield  {author} {\bibinfo {author} {\bibfnamefont {U.~A.}\ \bibnamefont
  {Wiedemann}}\ and\ \bibinfo {author} {\bibfnamefont {U.}~\bibnamefont
  {Heinz}},\ }\href {\doibase http://dx.doi.org/10.1016/S0370-1573(99)00032-0}
  {\bibfield  {journal} {\bibinfo  {journal} {Phys. Rep.}\ }\textbf {\bibinfo
  {volume} {319}},\ \bibinfo {pages} {145} (\bibinfo {year}
  {1999})}\BibitemShut {NoStop}%
\bibitem [{\citenamefont {Heinz}\ and\ \citenamefont {Jacak}(1999)}]{hbt2}%
  \BibitemOpen
  \bibfield  {author} {\bibinfo {author} {\bibfnamefont {U.}~\bibnamefont
  {Heinz}}\ and\ \bibinfo {author} {\bibfnamefont {B.~V.}\ \bibnamefont
  {Jacak}},\ }\href@noop {} {\bibfield  {journal} {\bibinfo  {journal} {Annu.
  Rev. Nucl. Part. Sci.}\ }\textbf {\bibinfo {volume} {49}},\ \bibinfo {pages}
  {529} (\bibinfo {year} {1999})}\BibitemShut {NoStop}%
\bibitem [{\citenamefont {Lisa}\ \emph {et~al.}(2005)\citenamefont {Lisa},
  \citenamefont {Pratt}, \citenamefont {Soltz},\ and\ \citenamefont
  {Wiedemann}}]{hbt3}%
  \BibitemOpen
  \bibfield  {author} {\bibinfo {author} {\bibfnamefont {M.~A.}\ \bibnamefont
  {Lisa}}, \bibinfo {author} {\bibfnamefont {S.}~\bibnamefont {Pratt}},
  \bibinfo {author} {\bibfnamefont {R.}~\bibnamefont {Soltz}}, \ and\ \bibinfo
  {author} {\bibfnamefont {U.}~\bibnamefont {Wiedemann}},\ }\href@noop {}
  {\bibfield  {journal} {\bibinfo  {journal} {Annu. Rev. Nucl. Part. Sci.}\
  }\textbf {\bibinfo {volume} {55}},\ \bibinfo {pages} {357} (\bibinfo {year}
  {2005})}\BibitemShut {NoStop}%
\bibitem [{\citenamefont {He}\ \emph {et~al.}(2016{\natexlab{b}})\citenamefont
  {He}, \citenamefont {Qian},\ and\ \citenamefont {Huo}}]{rhhe20162}%
  \BibitemOpen
  \bibfield  {author} {\bibinfo {author} {\bibfnamefont {R.}~\bibnamefont
  {He}}, \bibinfo {author} {\bibfnamefont {J.}~\bibnamefont {Qian}}, \ and\
  \bibinfo {author} {\bibfnamefont {L.}~\bibnamefont {Huo}},\ }\href
  {http://link.aps.org/doi/10.1103/PhysRevC.94.034902} {\bibfield  {journal}
  {\bibinfo  {journal} {Phys. Rev. C}\ }\textbf {\bibinfo {volume} {94}},\
  \bibinfo {pages} {034902} (\bibinfo {year} {2016}{\natexlab{b}})}\BibitemShut
  {NoStop}%
\bibitem [{\citenamefont {De}\ \emph {et~al.}(2013)\citenamefont {De},
  \citenamefont {Tarnowsky}, \citenamefont {Nayak}, \citenamefont
  {Scharenberg},\ and\ \citenamefont {Srivastava}}]{2013profile}%
  \BibitemOpen
  \bibfield  {author} {\bibinfo {author} {\bibfnamefont {S.}~\bibnamefont
  {De}}, \bibinfo {author} {\bibfnamefont {T.}~\bibnamefont {Tarnowsky}},
  \bibinfo {author} {\bibfnamefont {T.~K.}\ \bibnamefont {Nayak}}, \bibinfo
  {author} {\bibfnamefont {R.~P.}\ \bibnamefont {Scharenberg}}, \ and\ \bibinfo
  {author} {\bibfnamefont {B.~K.}\ \bibnamefont {Srivastava}},\ }\href@noop {}
  {\bibfield  {journal} {\bibinfo  {journal} {Phys. Rev. C}\ }\textbf {\bibinfo
  {volume} {88}},\ \bibinfo {pages} {044903} (\bibinfo {year}
  {2013})}\BibitemShut {NoStop}%
\bibitem [{\citenamefont {Zhang}\ \emph {et~al.}(2004)\citenamefont {Zhang},
  \citenamefont {Efaaf},\ and\ \citenamefont {Wong}}]{WNZhangGran1}%
  \BibitemOpen
  \bibfield  {author} {\bibinfo {author} {\bibfnamefont {W.~N.}\ \bibnamefont
  {Zhang}}, \bibinfo {author} {\bibfnamefont {M.~J.}\ \bibnamefont {Efaaf}}, \
  and\ \bibinfo {author} {\bibfnamefont {C.-Y.}\ \bibnamefont {Wong}},\ }\href
  {http://link.aps.org/doi/10.1103/PhysRevC.70.024903} {\bibfield  {journal}
  {\bibinfo  {journal} {Phys. Rev. C}\ }\textbf {\bibinfo {volume} {70}},\
  \bibinfo {pages} {024903} (\bibinfo {year} {2004})}\BibitemShut {NoStop}%
\bibitem [{\citenamefont {Zhang}\ \emph {et~al.}(2006)\citenamefont {Zhang},
  \citenamefont {Ren},\ and\ \citenamefont {Wong}}]{WNZhangGran2}%
  \BibitemOpen
  \bibfield  {author} {\bibinfo {author} {\bibfnamefont {W.-N.}\ \bibnamefont
  {Zhang}}, \bibinfo {author} {\bibfnamefont {Y.-Y.}\ \bibnamefont {Ren}}, \
  and\ \bibinfo {author} {\bibfnamefont {C.-Y.}\ \bibnamefont {Wong}},\ }\href
  {http://link.aps.org/doi/10.1103/PhysRevC.74.024908} {\bibfield  {journal}
  {\bibinfo  {journal} {Phys. Rev. C}\ }\textbf {\bibinfo {volume} {74}},\
  \bibinfo {pages} {024908} (\bibinfo {year} {2006})}\BibitemShut {NoStop}%
\bibitem [{\citenamefont {Zhang}\ \emph {et~al.}(2009)\citenamefont {Zhang},
  \citenamefont {Yang},\ and\ \citenamefont {Ren}}]{WNZhangGran3}%
  \BibitemOpen
  \bibfield  {author} {\bibinfo {author} {\bibfnamefont {W.-N.}\ \bibnamefont
  {Zhang}}, \bibinfo {author} {\bibfnamefont {Z.-T.}\ \bibnamefont {Yang}}, \
  and\ \bibinfo {author} {\bibfnamefont {Y.-Y.}\ \bibnamefont {Ren}},\ }\href
  {http://link.aps.org/doi/10.1103/PhysRevC.80.044908} {\bibfield  {journal}
  {\bibinfo  {journal} {Phys. Rev. C}\ }\textbf {\bibinfo {volume} {80}},\
  \bibinfo {pages} {044908} (\bibinfo {year} {2009})}\BibitemShut {NoStop}%
\bibitem [{\citenamefont {Hu}\ \emph {et~al.}(2015)\citenamefont {Hu},
  \citenamefont {Zhang},\ and\ \citenamefont {Ren}}]{WNZhangGran4}%
  \BibitemOpen
  \bibfield  {author} {\bibinfo {author} {\bibfnamefont {Y.}~\bibnamefont
  {Hu}}, \bibinfo {author} {\bibfnamefont {W.-N.}\ \bibnamefont {Zhang}}, \
  and\ \bibinfo {author} {\bibfnamefont {Y.-Y.}\ \bibnamefont {Ren}},\ }\href
  {http://stacks.iop.org/0954-3899/42/i=4/a=045105} {\bibfield  {journal}
  {\bibinfo  {journal} {J. Phys. G: Nucl. Part. Phys.}\ }\textbf {\bibinfo
  {volume} {42}},\ \bibinfo {pages} {045105} (\bibinfo {year}
  {2015})}\BibitemShut {NoStop}%
\bibitem [{\citenamefont {Pratt}(1994)}]{paper9added1}%
  \BibitemOpen
  \bibfield  {author} {\bibinfo {author} {\bibfnamefont {S.}~\bibnamefont
  {Pratt}},\ }\href {http://link.aps.org/doi/10.1103/PhysRevC.49.2722}
  {\bibfield  {journal} {\bibinfo  {journal} {Phys. Rev. C}\ }\textbf {\bibinfo
  {volume} {49}},\ \bibinfo {pages} {2722} (\bibinfo {year}
  {1994})}\BibitemShut {NoStop}%
\bibitem [{\citenamefont {Randrup}(2005)}]{paper9added2}%
  \BibitemOpen
  \bibfield  {author} {\bibinfo {author} {\bibfnamefont {J.}~\bibnamefont
  {Randrup}},\ }\href {\doibase 10.1556/aph.22.2005.1-2.8} {\bibfield
  {journal} {\bibinfo  {journal} {Acta Physica Hungarica Series A, Heavy Ion
  Physics}\ }\textbf {\bibinfo {volume} {22}},\ \bibinfo {pages} {69} (\bibinfo
  {year} {2005})}\BibitemShut {NoStop}%
\bibitem [{\citenamefont {Schulc}\ and\ \citenamefont
  {Tomášik}(2010)}]{paper9added3}%
  \BibitemOpen
  \bibfield  {author} {\bibinfo {author} {\bibfnamefont {M.}~\bibnamefont
  {Schulc}}\ and\ \bibinfo {author} {\bibfnamefont {B.}~\bibnamefont
  {Tomášik}},\ }\href {\doibase 10.1140/epja/i2010-10984-0} {\bibfield
  {journal} {\bibinfo  {journal} {Eur. Phys. J. A}\ }\textbf {\bibinfo {volume}
  {45}},\ \bibinfo {pages} {91} (\bibinfo {year} {2010})}\BibitemShut {NoStop}%
\end{thebibliography}
\end{document}